\documentclass[onecolumn]{aa}
\usepackage{graphicx}
\usepackage{txfonts}
\usepackage{natbib}
\bibpunct{(}{)}{;}{a}{}{,} 
\begin{document}
   \title{Optical Nuclei of Radio-loud AGN and the Fanaroff-Riley Divide}
   \author{P. Kharb 
          \and
          P. Shastri
          }
   \offprints{P. Kharb, \email{rhea@iiap.res.in}}

   \institute{Indian Institute of Astrophysics, Koramangala,
              Bangalore - 560 034, India}

\titlerunning{Optical nuclei and the F-R Divide}
\authorrunning{P. Kharb \& P. Shastri}


\abstract{We investigate the nature of the point-like optical 
nuclei that have been found in the centres of the host galaxies 
of a majority of radio galaxies by the {\it Hubble Space Telescope}. 
We examine the evidence that these optical nuclei are relativistically 
beamed, and look for differences in the behaviour of the nuclei found 
in radio galaxies of the two Fanaroff-Riley types. We also attempt to 
relate this behaviour to the properties of the optical nuclei in their 
highly beamed counterparts (the BL~Lac objects and radio-loud quasars) 
as hypothesized by the simple Unified Scheme. 
Simple model-fitting of the data suggests that the emission may be coming 
from a non-thermal relativistic jet. It is also suggestive that the 
contribution from an accretion disk is not significant for the 
FRI objects and for the narrow-line radio galaxies of FRII type, 
while it may be significant for the Broad-line objects, and 
consistent with the idea that the FRII optical nuclei seem to suffer 
from extinction due to an obscuring torus while the FRI optical nuclei
do not. These results are broadly in agreement with the Unified Scheme for 
radio-loud AGNs.

   \keywords{galaxies: active -- BL~Lacertae objects: general --
galaxies: nuclei -- quasars: general
               }
   }

   \maketitle

\section{Introduction}
\label{intro}

The `radio-loud' active galactic nuclei (AGNs) which include radio galaxies, 
BL~Lac objects and quasars, show twin lobes of synchrotron-emitting
plasma connected to a `core' by plasma jets on scales of $\sim$~100~kpc. 
\citet{FanaroffRiley74} recognised that the radio morphology of radio 
galaxies along with their total radio power (at 178~MHz) fell into 
two distinct 
subclasses: the lower-power Fanaroff-Riley type I (FRI) objects show 
extended plumes and tails with no distinct termination of the jet while 
the higher-power type II (FRII) objects show narrow, collimated jets 
and terminal `hotspots'. 
The FRII radio galaxies have systematically more luminous 
optical emission lines \citep{ZirbelBaum95} while FRI radio galaxies 
inhabit richer environments \citep{Prestage88}; the value of the FRI/FRII 
radio luminosity break blurs at higher radio frequencies 
\citep[see][]{UrryPadovani95}, and increases with the optical 
luminosity of the host galaxy \citep{LedlowOwen96}. 
The origin of the F--R dichotomy is far from 
clear: suggested possibilities include differences in the spin of 
the supermassive black hole resulting in different jet kinetic powers 
\citep{Baum95,Meier99}, galaxy environments \citep{SmithHeckman90}, and 
accretion rates \citep{Baum95}. 
The dichotomy issue is complicated by the observations of
sources having both FRI and FRII characteristics 
\citep[FRI/II, e.g.,][]{Capetti95}.

AGN jets experience bulk relativistic motion \citep{BlandfordKonigl79}
resulting in orientation playing a dominant role in their appearance 
and a simple Unified Scheme (US) has emerged 
\citep[e.g.,][and refs. therein]{UrryPadovani95} according to which the 
BL~Lac objects and radio-loud quasars are the relativistically beamed 
counterparts of FRI and FRII radio galaxies, respectively. 
Apart from bulk relativistic motion in radio galaxies,
the US requires a ubiquitous optically thick torus in the FRII class 
of objects 
($i.e.,$ FRII radio galaxies and radio-loud quasars)
which  hides the powerful optical continuum and broad 
emission lines from the 
nucleus in the edge-on objects, while no such torus is required by the US 
for the FRI class of objects as broad emission lines are weak/absent, 
and it has not been clear whether or not a torus exists.

The beamed synchrotron emission from the base of the jet or `core' 
must extend to visible wavelengths and there is strong evidence for it
in BL~Lacs and quasars \citep{Impey90,Wills92}. Recently, evidence for 
an optical synchrotron component in the relatively unbeamed radio galaxies 
has also surfaced, in the form of unresolved nuclear sources in the high 
resolution images with the {\it Hubble Space Telescope} ($HST$) 
\citep[e.g.,][]{Chiaberge99,Chiaberge02,Hardcastle00,Verdoes02}. 
These authors argue on the basis of the strong connection with the radio 
core emission, anisotropy \citep{Capetti99} and colour information
that these optical nuclei are indeed due to synchrotron radiation.

In this paper, we further test the idea that the unresolved nuclear
optical emission from radio galaxies is beamed synchrotron emission from
the base of the jet, using the radio core prominence parameter ($R_{c}$)
as an indicator of the orientation of the AGN axes. 
We then attempt to place 
these correlations in the broader framework of the US and 
test for consistencies. 
We come up with a model-fitting approach to investigate quantitatively 
the dependence of the optical emission on orientation and further test the 
predictions of the US in terms of the presence (or absence) of 
obscuring tori and the contribution of thermal accretion disks.  
We list the caveats with regard to our current sample and attempt 
to address them.
The outline of the paper is as follows : in Sect.~2 we 
discuss the optical 
nuclei in FRI and FRII radio galaxies and the correlations with $R_{c}$. 
In Sect.~3 we compare the optical nuclei with those in BL~Lacs 
and quasars 
and discuss the results along with model-fitting. The
model equations and the fitting procedure are described in 
Appendix~\ref{appmodel} and \ref{appfit}. Sect.~4 lists the 
conclusions. Throughout this paper, 
$H_0$~=~75~km~$s^{-1}$~Mpc$^{-1}$ and $q_0~=~0.5$ have been adopted and 
the spectral index $\alpha$ is defined such that $F_{\nu}=\nu^{-\alpha}$.

\section{The optical nuclei in FRI and FRII radio galaxies}

Optical nuclei have been detected in a majority of 3CR, B2 and UGC FRI and 
FRII radio galaxies with the WFPC2 on board the {\it HST} which appear as 
unresolved sources with angular sizes $\sim$~0$\arcsec$.1. 
The results of studies based on this discovery have been 
presented by \citet{Chiaberge99,Capetti99,Hardcastle00,
Capetti02,Chiaberge02} and \citet{Verdoes02}.

For our study, we chose an eclectic sample of FRI and FRII radio 
galaxies with either such a detected optical nucleus or with an upper limit 
to its optical flux density from the above-mentioned papers.  
Our set of FRI radio galaxies comprise 25 3CR \citep{Chiaberge99}, 
17 B2 \citep{Capetti02} and 10 UGC FRIs \citep{Verdoes02} 
along with NGC~7052 and NGC~6251 from \citet{Capetti99} 
and \citet{Hardcastle99} respectively. Objects with ambiguous 
morphologies (e.g., FRI/II sources mentioned in Sect.~\ref{intro})
are excluded. So is 3C~386 whose optical ``nucleus'' is 
in fact a foreground star \citep{Chiaberge02}. We thus have 54 FRI radio 
galaxies spanning a redshift range of 0.0037~$\leq z \leq$~0.29. 
The FRII radio galaxies include 53 objects from the 3CR sample 
presented in \citet{Chiaberge02} and 2 B2 FRIIs from \citet{Capetti02}. 
Among the 55 FRIIs considered, 
there are 42 narrow-line radio galaxies (NLRGs) 
and 13 broad-line radio galaxies (BLRGs). The FRII radio galaxies span a 
redshift range of 0.025~$\leq z \leq$~0.296. 

Tables \ref{tb-FRI} and \ref{tb-FRII} list the FRI and FRII radio galaxies 
respectively, along with their optical and radio data. 
Col.~1 lists the IAU name; 
Col.~2: alternative name; 
Col.~3: redshift (from the references for radio core data,
except for UGC FRIs which are from NED); 
Col.~4: dust disk minor-to-major axis ratio (superscripts `$d$' and `$l$' 
stand for disk and lane respectively) 
from \citet{Verdoes99} except 3C~83.1, 3C~296, 
3C~449, 3C~465, 3C~326 and 3C~452 which are from \citet{deKoff00};
Col.~5: logarithm of extended radio luminosity at 1.4 GHz
in W Hz$^{-1}$, calculated using the difference between total and core 
flux density; data at 5 GHz were converted to 1.4 GHz using 
$\alpha_{radio}^{ext}$ = 0.7 for extended radio emission;
Col.~6: 5 GHz radio core flux density in mJy; 
Col.~7: reference for the radio core (and total flux density if different);
Col.~8: logarithm of radio core prominence standardized to an emitted 
wavelength of 6~cm;
Col.~9: nuclear optical luminosity in W Hz$^{-1}$ estimated at an 
emitted wavelength of 5500 $\AA$;
Col.~10: reference for nuclear optical flux density/luminosity.

\subsection{The correlations with radio core prominence for radio galaxies}
\label{secRGs}

The radio core prominence parameter, which is the ratio of the 
core-to-extended radio flux density 
($R_{c} \equiv S_{core}/S_{ext}$) is 
a known statistical indicator of orientation \citep{KapahiSaikia82,
OrrBrowne82} assuming that the core is the unresolved relativistically 
beamed nuclear jet and the lobes are unbeamed. $R_{c}$ has indeed been 
shown to correlate with other orientation-dependent properties both in FRIIs 
\citep[e.g.,][]{KapahiSaikia82} and FRIs \citep[e.g.,][]{Laing99}. 
We use the parameter $R_{c}$ to test if the luminosities of the 
optical nuclei $L_o$, are
orientation-dependent. If the intrinsic optical synchrotron emission from 
the jet is relativistically beamed by the Doppler factor 
$\delta$ where $\delta \equiv [\gamma(1-\beta$ cos$\theta)]^{-1}$, 
$\gamma$ being the Lorentz factor ($\gamma \equiv 1/\sqrt{1-\beta^2}$),
$\beta$ the bulk velocity in units of the speed of light, and
$\theta$ being the angle between the radio axis and our line of sight, 
then $L_o$ should correlate with $R_{c}$. We note that $L_o$ has been shown 
to correlate with the radio core {\it luminosity} 
by \citet{Chiaberge99,Hardcastle00}.

\begin{figure}[h]
\centerline{
\includegraphics[height=6.5cm]{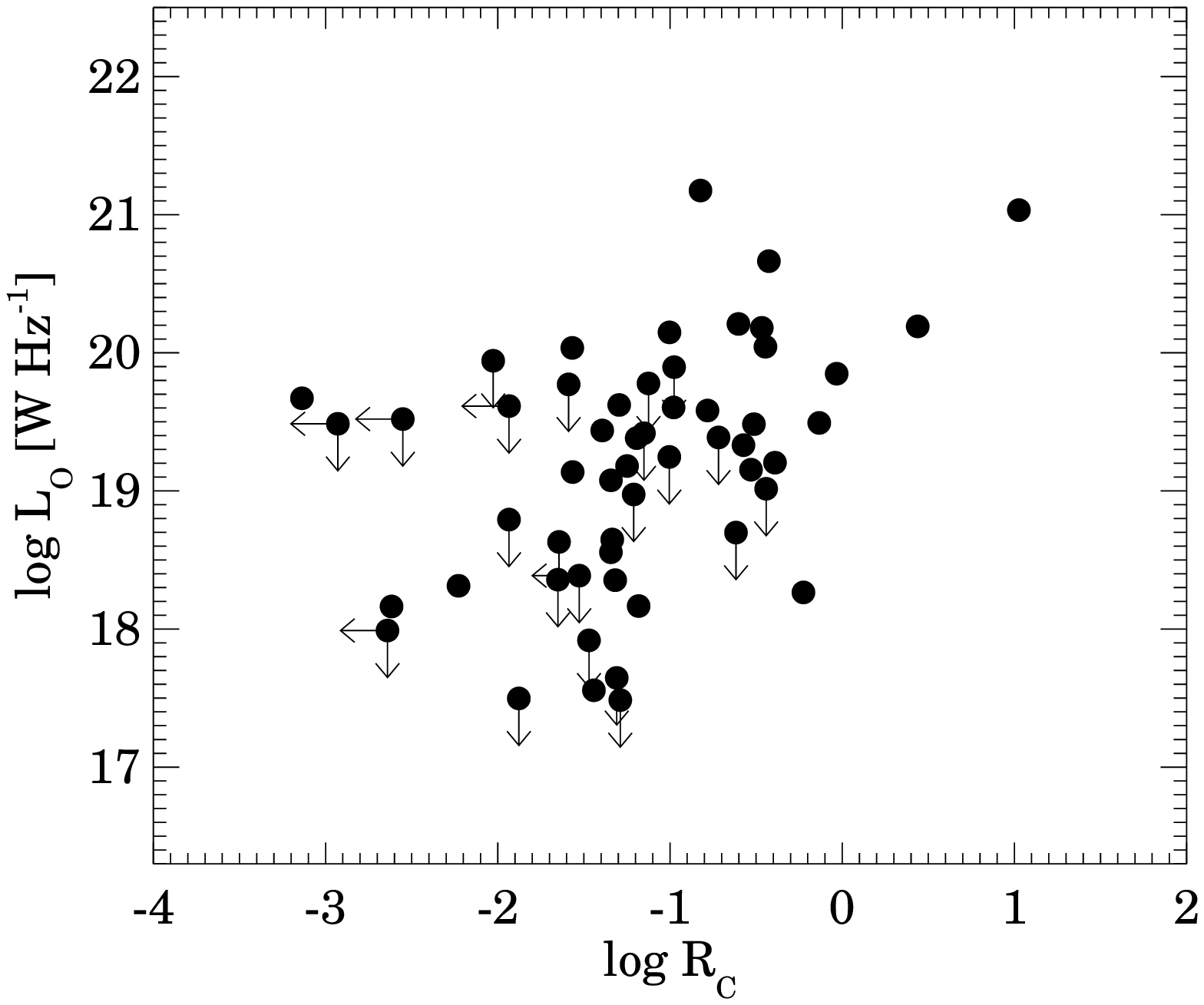}
\includegraphics[height=6.5cm]{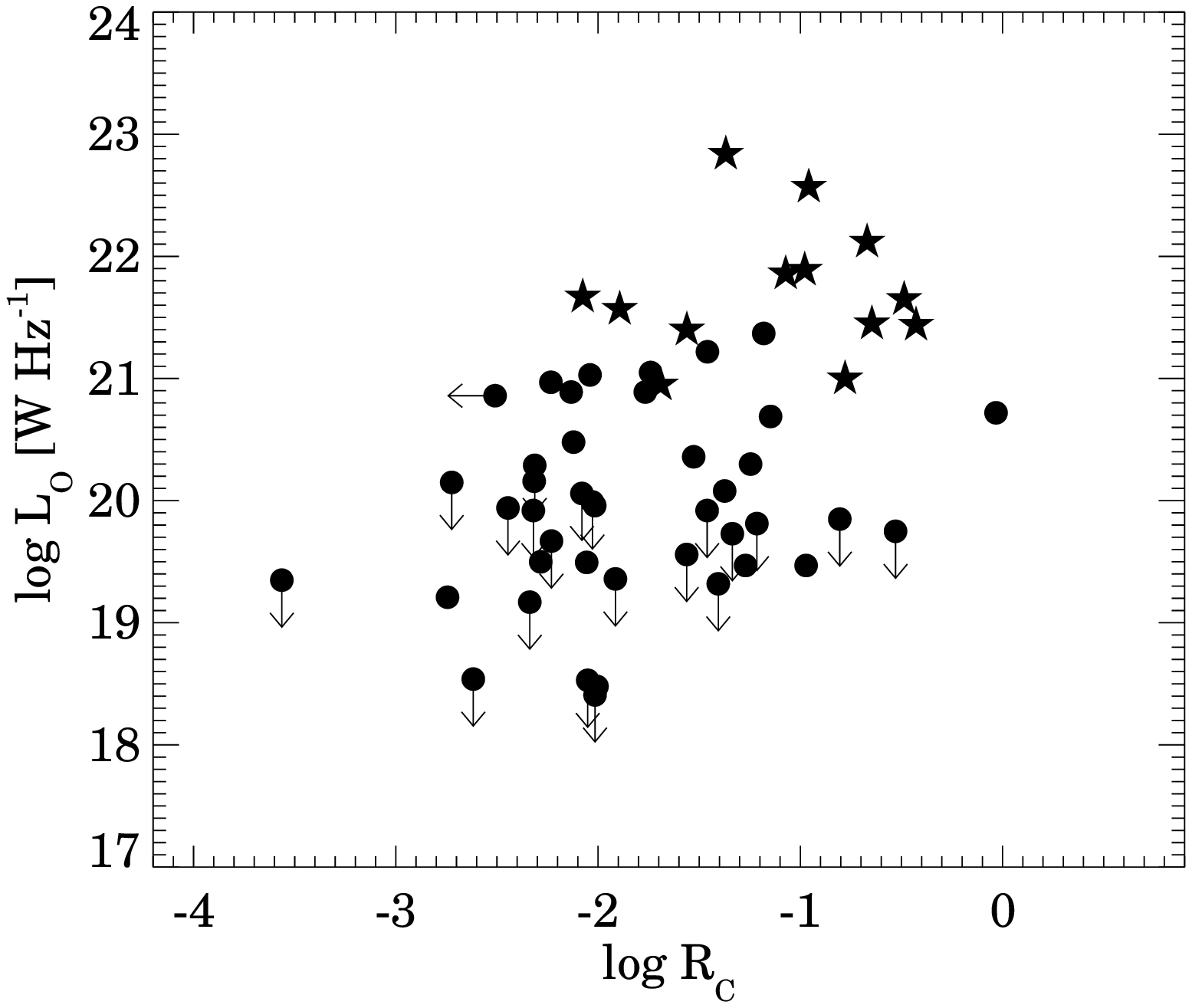}}
\caption{The luminosity of the optical nuclei $L_{o}$, plotted against the 
radio core prominence $R_{c}$, for the FRI (left) and FRII
radio galaxies (right). $\bullet$ radio galaxies, 
$\star$ BLRGs, $\downarrow$ and $\leftarrow$ 
upper limits. Statistics for the fits are listed in Table~\ref{tb-ASURV}.}
\label{lo-Rc-FRI/II}
\end{figure}

The optical luminosities of the unresolved {\it HST} nuclei were 
{\it K}-corrected and calculated at an emitted wavelength of 5500~$\AA$, 
assuming an optical spectral index $\alpha_{opt}$~=~1. 
$R_{c}$ was calculated using observed radio core and 
total flux densities at 5~GHz and was further {\it K}-corrected to an 
emitted frequency of 5~GHz. For some sources flux densities were estimated 
from 1.4~GHz assuming $\alpha_{radio}^{ext}$~=~0.7 and 
$\alpha_{radio}^{core}$~=~0 for the extended and core radio emission, 
respectively. In Fig.~\ref{lo-Rc-FRI/II} we plot $L_o$ versus 
$R_{c}$ for the FRI and FRII radio galaxies. 

We note that there are many upper limits to $L_o$. 
We have analysed the statistical significance of the correlations with 
the aid of the Astronomical Survival Analysis (ASURV) package as 
implemented in IRAF, which takes into account data which are only 
upper/lower limits. $L_{o}$ turns out to be significantly correlated 
with $R_{c}$ for the FRI radio galaxies ({\it p} = 0.0001, generalized
Spearman Rank test, see Table~\ref{tb-ASURV}), arguing that the 
nuclear optical 
emission is orientation-dependent in the same sense as the radio 
emission and may also originate in the relativistically beamed jet.
The implication of the above result is 
consistent with what \citet{Verdoes02} suggest for their UGC FRI sample, 
{\it viz.,} that beaming also plays a role in the variance of $L_{o}$,
in addition to the intrinsic variance in 
the nuclear jet $L_{jet}^{int}$ which presumably ionizes the line-emitting 
gas.

For the FRII radio galaxies, the correlation is significant 
{\it only if the BLRGs 
(plotted as stars in Fig.~\ref{lo-Rc-FRI/II}) are included}, 
while the narrow-line 
objects do not show a significant correlation by themselves 
({\it p} $>0.2$, generalized Spearman Rank test). 
The narrow-line FRII galaxies show no correlation even with the more 
sensitive parametric Pearson's correlation test which however uses 
uncensored data ({\it p} $>0.1$). 
This lack of correlation could be explained by the 
presence of a dusty obscuring torus in FRII radio galaxies
that is hypothesized by the US; this could also result in the large number of  
non-detections \citep[also see][]{Chiaberge02}.

\subsection{Kpc-scale dust disks in FRI radio galaxies}
\label{secdust}

\begin{figure}[h]
\centerline{
\includegraphics[height=6.5cm]{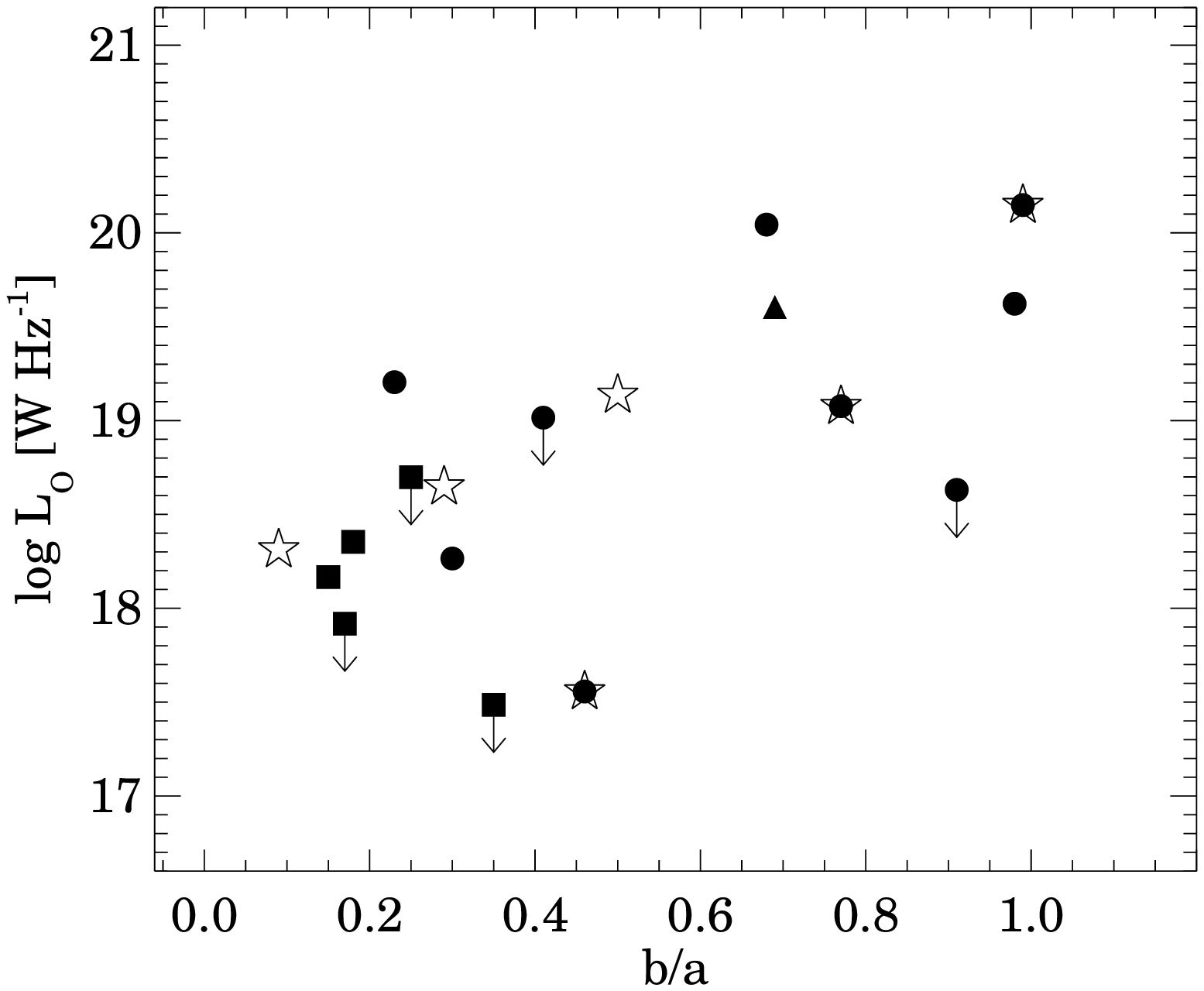}
\includegraphics[height=6.5cm]{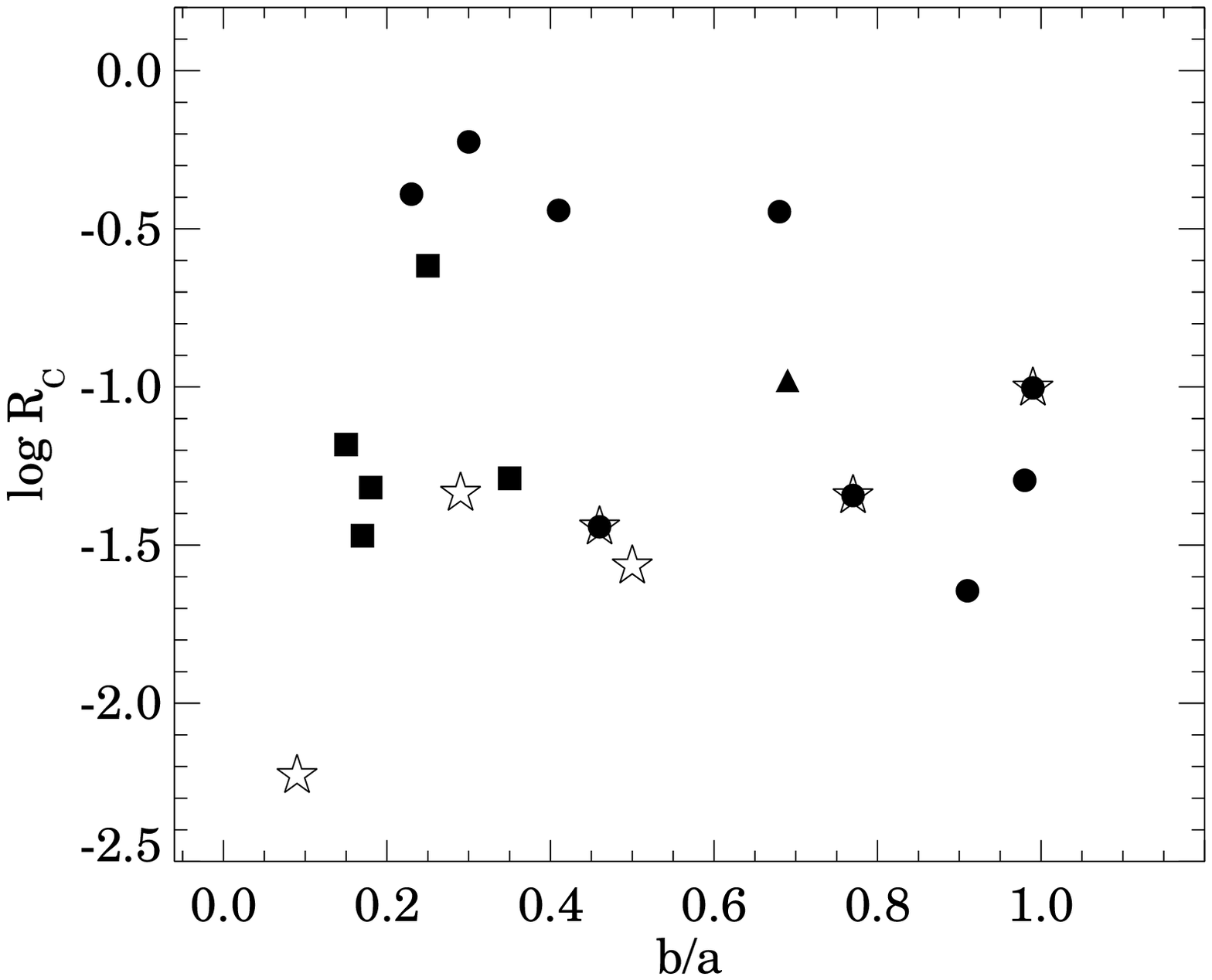}}
\caption{Nuclear optical luminosity $L_{o}$ (left) and radio core 
prominence $R_{c}$ (right) plotted against minor-to-major axis ratio $b/a$,
for the FRI radio galaxies. 
$\bullet$ and {\tiny $\blacksquare$} denote the dust disks and lanes 
respectively, from \citet{Verdoes99} while the open stars and 
$\blacktriangle$ denote the dust disks and lanes from \citet{deKoff00}, 
$\downarrow$ upper limits. The sources common to the two 
papers are 
shown by an open star superimposed by $\bullet$. Table~\ref{tb-ASURV} 
lists the statistics for the correlations.}
\label{lo-Rc-ba-FRI}
\end{figure}

While the evidence for an obscuring torus in FRIs is so far meagre, 
much larger dust disks and lanes of sizes $\sim$ 100 pc to a 
few kpc have been discovered in many FRI radio galaxies 
\citep[e.g.,][]{Verdoes99,deKoff00}. It has been suggested by 
\citet{Verdoes99,Capetti99} and \citet{deKoff00} that the kpc-scale radio 
jet tends to align with the axis of this disk. We investigate this 
point here for the subset of objects where data on dusty disks, as well 
as $L_{o}$ and $R_c$ are available. We find different relations of the 
minor-to-major axis ratio $b/a$, of the extended dust disk with $L_{o}$ and 
$R_c$ for the samples presented in the above papers 
(see Table~\ref{tb-ASURV}).
$b/a$ of the dust disks correlates significantly both with the 
$L_{o}$ and $R_c$ for the \citeauthor{deKoff00} FRI galaxies.  
$b/a$ correlates significantly with $L_{o}$ for the \citeauthor{Verdoes99} 
FRI sources {\it only when both} disks and lanes are considered together. 
However, they show no correlation with $R_c$ when disks and lanes are taken 
together, and a correlation in the opposite sense to that predicted, when 
only disks are considered. When the objects from both samples are combined
and both dust disks and lanes are considered, $b/a$ correlates with $L_{o}$
but not with $R_c$ (see Fig.~\ref{lo-Rc-ba-FRI}, and 
Table~\ref{tb-ASURV} for the statistical results).

It thus appears that the axes of the extended dust disks do not tend to 
be aligned with the orientation of the AGN, but that these disks could 
be causing some extinction of the optical nuclear emission in 
FRI radio galaxies. This extinction would of course contribute to the 
scatter in the $L_{o}$ -- $R_c$ correlation.

\section{Comparison with the optical nuclei of the beamed objects and the US}

In the simple US, the beamed counterparts of the FRI and FRII 
radio galaxies are the BL~Lac objects and the radio-loud quasars 
respectively. 
In Sect.~\ref{secRGs} we find the optical emission from 
galaxy nuclei to be 
orientation-dependent. Given that optical emission from BL~Lacs
and quasars is also beamed \citep[e.g.,][]{KapahiShastri87,Baker94}, we 
attempt to relate the behaviour of the galaxy nuclei to that of 
BL~Lacs and quasars in the framework of the US. We use this framework 
to extend 
the correlations of $L_{o}$ with $R_{c}$ to higher values 
of $R_{c}$. We consider the 
FRI radio galaxies and BL~Lacs together (the ``FRI population"), and 
similarly consider the FRII radio galaxies and radio-loud quasars together 
(the ``FRII population"). To investigate 
quantitatively the dependence of the optical emission on orientation,
we come up with a model-fitting approach. We attempt to apply this to the 
available data and present the results in Sect.~\ref{secresults}. 
We further outline the caveats and the drawbacks of our current sample 
and attempt to address them in Sect.~\ref{secmatched}.

\subsection{The data}

The set of BL~Lac objects we considered comprise 
both radio-selected and X-ray selected BL~Lacs 
from \citet{PerlmanStocke93,VermeulenCohen94} and 
\citet{Laurent-Muehleisen93}, thus including objects having both high and 
intermediate $R_c$ values. 
After excluding BL~Lacs which showed FRII 
radio morphology in the form of terminal hotspots $viz.,$ 1308+326, 
1823+568, 2007+777 \citep{Kollgaard92}, 1749+701 \citep{O'Dea88} and 1803+784 
\citep{Cassaro99}; which were gravitational microlensing candidates, 
$viz.,$ 1413+135 
\citep[this object also has other peculiarities like a spiral 
host galaxy;][]{PerlmanStocke93} and which had uncertain redshifts 
$viz.,$ 0716+714, our
BL~Lac sample consists of 44 objects spanning a redshift range 
of 0.028~$\leq z \leq$~0.997. 
We have considered 34 high $R_c$ radio-loud quasars 
from \citet{VermeulenCohen94} spanning the redshift range of 
0.158~$\leq z \leq$~2.367. 

We have taken the total optical luminosity of BL~Lacs and quasars (as 
derived from their available {\it V}-band magnitudes) as the nuclear optical 
luminosity, {\it assuming that the nucleus overwhelms the host galaxy 
emission}. As the BL~Lacs are known to be strongly variable, we 
took radio and optical measurements from the literature that were as 
closely spaced in time as were available. Several {\it V}-magnitudes come 
from optical monitoring campaigns of \citet{Pica88,Webb88} and 
\citet{Falomo94}. Quasars can also be optically violent variables (OVVs)
but they constitute less than 25$\%$ of our quasar sample. 
The data are tabulated in Tables~\ref{tb-BLL} and \ref{tb-QSR}. 
Col.~1 lists the IAU name (B1950);
Col.~2: alternative name;
Col.~3: redshift (from the references for radio core data, except 1402+042,
0333+321, 0835+580 and 0836+710 which are from \citet{Veron98});
Col.~4: {\it V}-band magnitude;
Col.~5: reference for $m_v$;
Col.~6: logarithm of extended radio luminosity at 1.4 GHz in 
W Hz$^{-1}$ -- taken from the reference for radio core flux density for 
BL~Lacs and calculated using core flux density and radio core prominence 
for quasars and the BL~Lacs 0454+844 and 0735+178; 
data at 5 GHz converted to 
1.4 GHz using $\alpha_{radio}^{ext}$ = 0.7 for the extended radio emission;
Col.~7: 5 GHz radio core flux density in mJy;
Col.~8: reference for the radio core and total flux density (for quasars
it is the reference for the radio core flux density and log$R_c$);
Col.~9: logarithm of radio core prominence standardized to an emitted 
wavelength of 6~cm;
Col.~10: nuclear optical luminosity in W Hz$^{-1}$ estimated at an 
emitted wavelength of 5500 $\AA$. The plots of $L_{o}$ against $R_{c}$ 
for the FRI and FRII populations are shown in Fig.~\ref{lo-rc-FRBQ}.

\subsection{Caveats}

While interpreting the $L_{o}$ -- $R_{c}$ plots, it is important to keep 
the following caveats in mind.\\
1.~~~The objects constitute an {\it eclectic} sample, with no rigorous 
selection criteria applied. \\
2.~~~The beamed and unbeamed objects 
are not matched in redshift, nor in extended radio luminosity. We discuss
the significance of this in Sect.~\ref{secpop} and try to define a 
`matched' sample in Sect.~\ref{secmatched}.\\
3.~~~The $L_{o}$ values for the BL~Lacs and quasars are derived from their 
total magnitudes, and include the host galaxy contribution.
Particularly in the intermediate $R_{c}$ regime for BL~Lacs, 
the host galaxies  
could contribute significantly to the assumed nuclear optical luminosity.\\
We address some of these issues later in the paper.

\subsection{Correlations with radio core prominence for the two populations}
\label{secpop}

For both the FR populations, the $L_{o}$ -- $R_{c}$ correlation $does$ 
extend to higher $R_{c}$, broadly consistent with the predictions of the US
and again reinforcing the idea that the optical nuclear emission is 
orientation-dependent in the same way as the radio core emission and it 
may thus constitute the optical counterpart of the relativistically beamed 
radio synchrotron jet.  
Using survival analysis, the generalized Spearman's Rank correlation 
test indicates that the FRI and FRII populations both show a significant 
correlation ({\it p} $<0.0001$) of $L_{o}$ with
$R_{c}$ (see Table~\ref{tb-ASURV}). 

\begin{figure}[h]
\centerline{
\includegraphics[height=6.5cm]{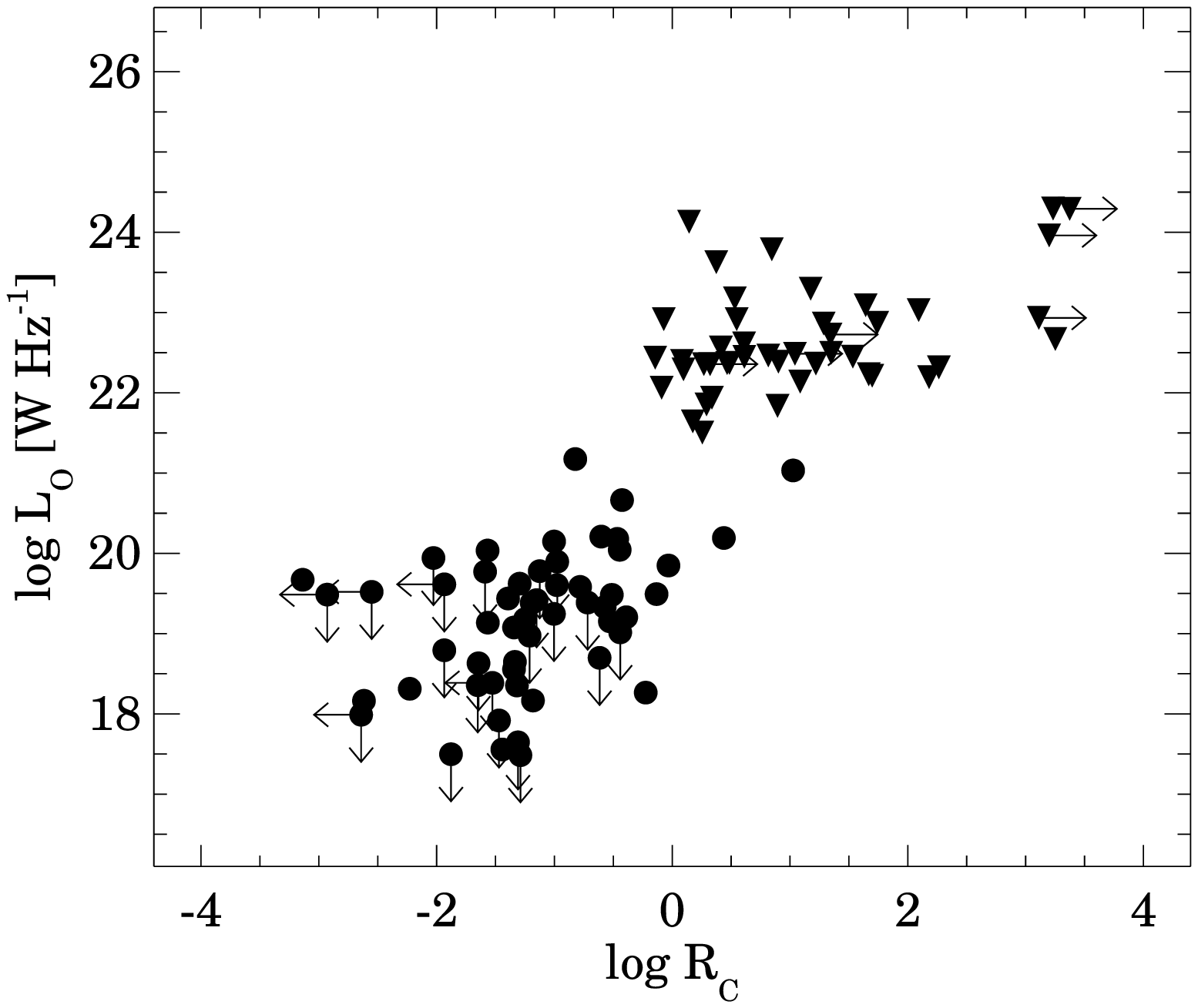}
\includegraphics[height=6.5cm]{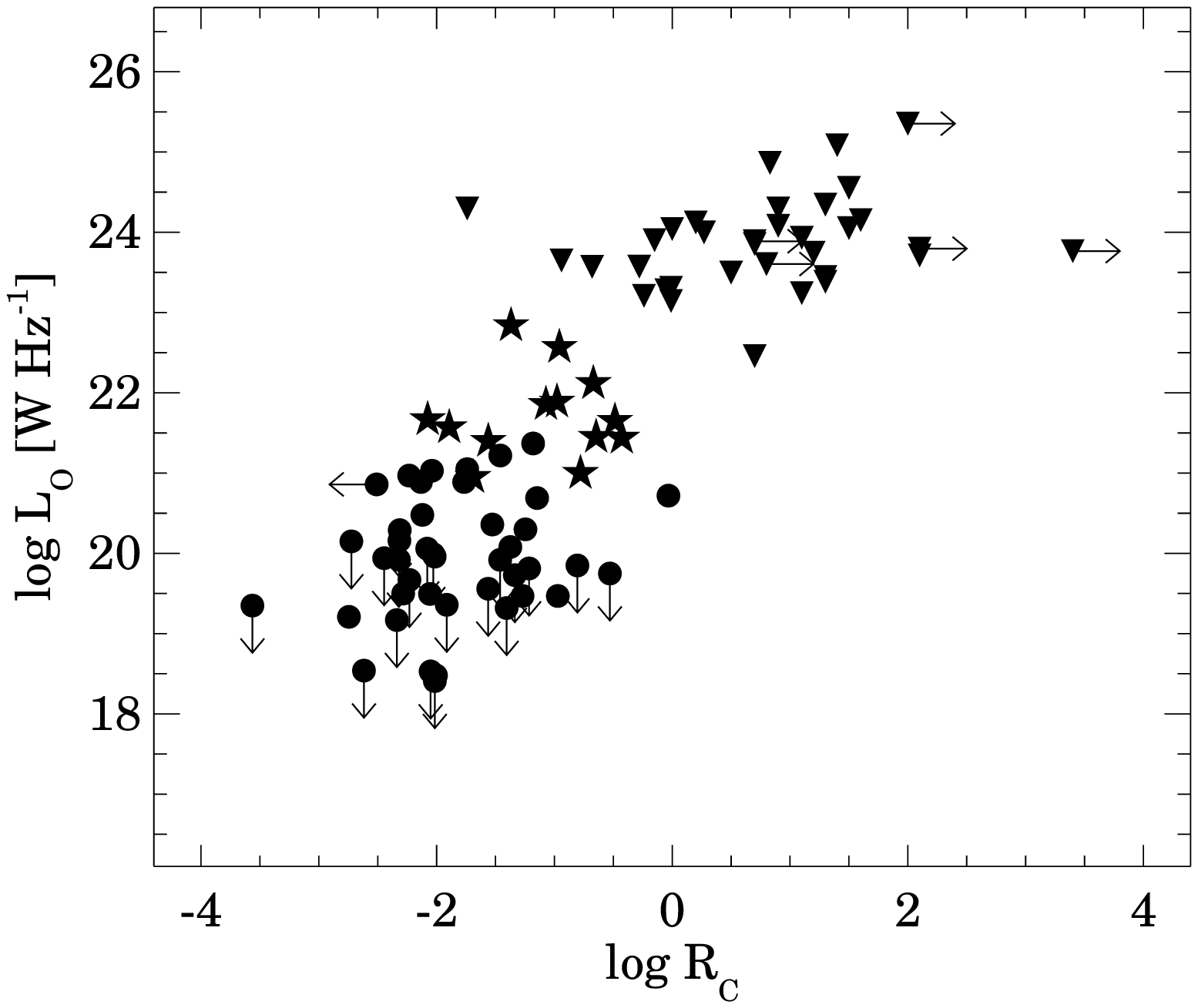}}
\caption{Nuclear optical luminosity $L_{o}$ versus radio core prominence $R_{c}$ 
for the FRI population (left) : $\bullet$ radio galaxies, 
{\tiny $\blacktriangledown$} BL~Lac objects. 
$L_{o}$ vs. $R_{c}$ for the FRII population (right) : $\bullet$ 
radio galaxies, {\tiny $\blacktriangledown$} radio-loud quasars, 
$\star$ BLRGs. $\downarrow$ and $\leftarrow$ upper 
limits, $\rightarrow$ lower limits. Statistics for the fits 
are listed in Table~\ref{tb-ASURV}.}
\label{lo-rc-FRBQ}
\end{figure}

For the FRII population, we showed in Sect.~\ref{secRGs} that the 
narrow-line FRIIs 
do not show any correlation by themselves. A significant correlation 
({\it p} $<0.0001$, generalized Spearman Rank test) 
exists for the broad-line 
objects, however, {\it i.e.,} for the the broad-line radio 
galaxies and quasars. 
These two observations taken together are consistent with there being 
obscuration effects by a torus in the FRIIs. Also, though fewer FRIIs 
than FRIs show 
detected optical nuclei, optical nuclei have been 
detected in {\it all} the BLRGs observed (where the US predicts no 
obscuration by the torus), again consistent with this idea. 
\citet{Chiaberge00} also suggest obscuration effects on the
basis of the non-detection of optical nuclei in some narrow-line FRIIs.

The BL~Lacs by themselves also show a significant correlation of 
$L_{o}$ with $R_c$, but the plot is flatter than what is expected 
from beaming alone. This could be due to the fact that their 
$L_{o}$ values include the contribution from the host galaxy, 
particularly since this contamination is likely to be more severe at 
intermediate values of $R_{c}$. Although for many BL~Lac objects
where host galaxies have been imaged \citep[e.g.,][]{Jannuzi97} 
the difference between the nuclear and the total optical luminosity ($L_{o}$) 
is less than the $50\%$ errors assumed in $L_{o}$ 
\citep[e.g.,][see Appendix~\ref{appfit}]{Verdoes02} 
for some sources this difference can as high as a magnitude 
\citep[e.g.,][]{Kotilainen98}. In principle, the flattening 
could also be due to the presence of a luminous accretion disk, in which 
case the BL~Lacs cannot be considered to be consistent with the unbeamed FRI
radio galaxies. The use of nuclear luminosities uncontaminated by host 
galaxy emission for {\it all} the objects would clarify the issue. We are 
in the process of investigating this point which is part of a future paper.

A two-dimensional Kolmogorov-Smirnov test shows that the FRI
and FRII populations are different at the $p < 0.0001$ level. For each of 
the populations, a multiple linear regression test using the statistics 
packages STATISTICA and ASURV (the `Buckley James' algorithm) of
$R_c$, redshift ($z$) and extended radio luminosity ($L_{ext}$) as independent 
variables, shows that the correlation coefficient for the $L_{o}$ -- $R_c$ 
correlation is the most significant ($p < 0.0001$). 
$L_{ext}$ is the next most significant contributor. Since $L_{ext}$ can 
reasonably be assumed to be an indicator of intrinsic AGN power, this implies 
that variation in intrinsic nuclear power contributes significantly to the 
scatter in the $L_{o}$ -- $R_c$ correlation.
As expected, the nuclear optical luminosity is correlated with redshift, both 
because luminosity is expected to correlate with redshift, and because of the 
absence of high redshift radio galaxies in the samples. 

\subsection{Matched subsamples of FRI and FRII objects}
\label{secmatched}

Ideally, all the objects in each population ought to be intrinsically 
similar in the framework of the US, which means that they 
should all be of similar intrinsic power, from the same volume of space, 
and with a narrow distribution of other orientation-independent parameters. 
As a next best step, we attempt here to derive a `matched' sample 
for the two FR populations, keeping in mind the multiple linear regression
results for the whole sample discussed in the previous section.

\begin{figure}[h]
\centerline{
\includegraphics[height=6.5cm]{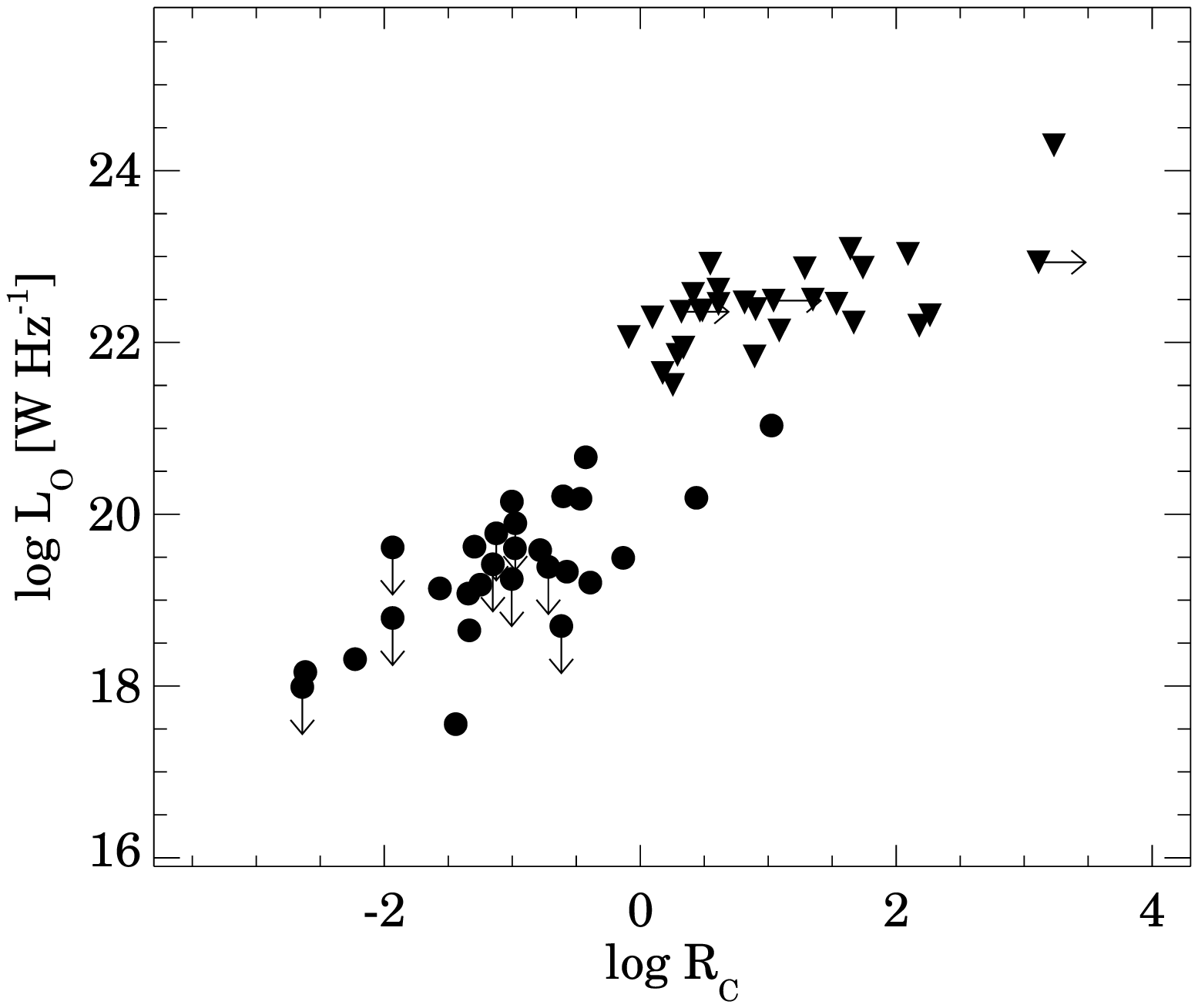}
\includegraphics[height=6.5cm]{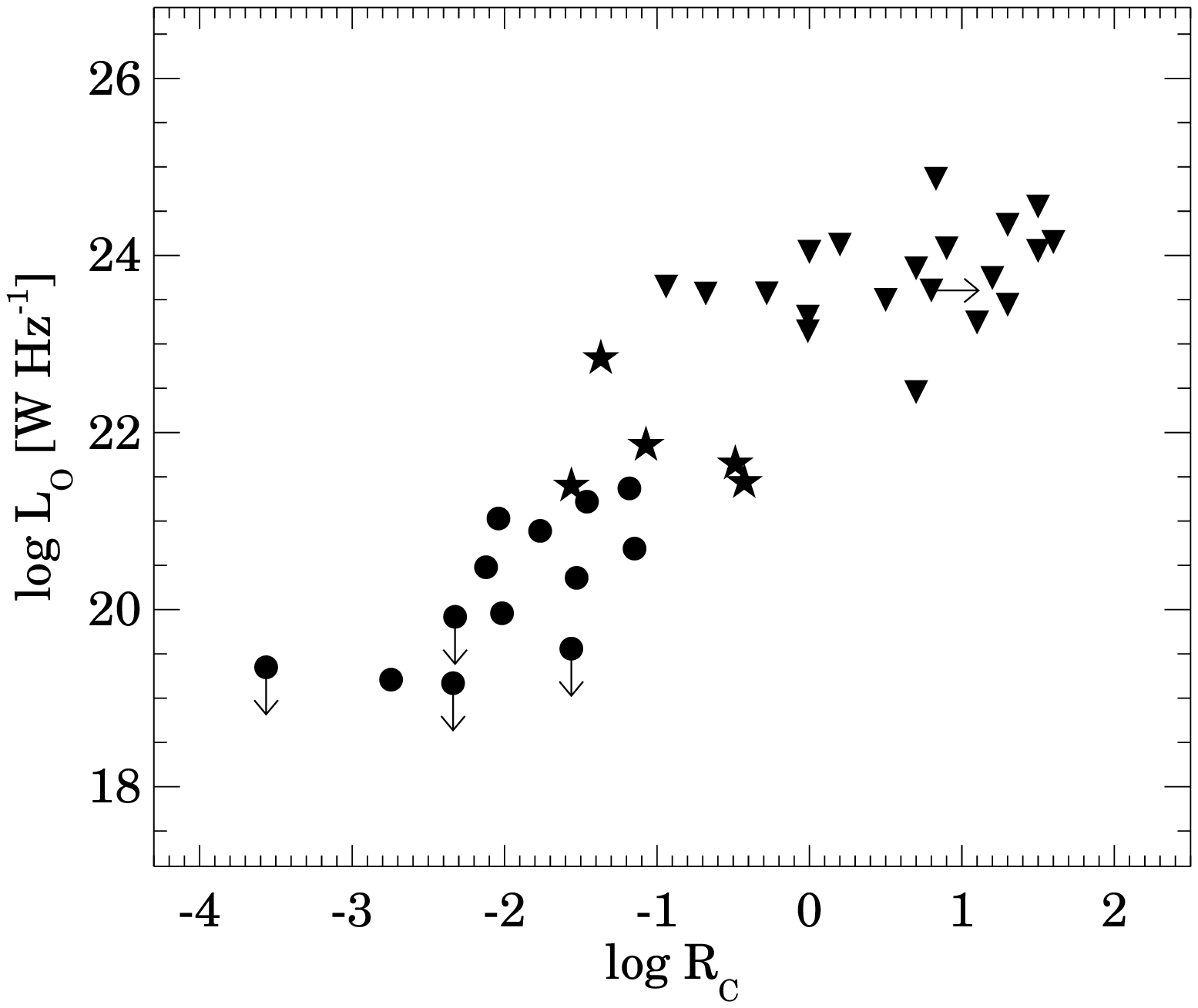}}
\caption{Nuclear optical luminosity $L_{o}$ versus radio core prominence
$R_{c}$ for the matched subsample of FRI galaxies and BL~Lac objects (left) 
and FRII galaxies and quasars (right). $\bullet$ radio galaxies, 
{\tiny $\blacktriangledown$} BL~Lac objects (left) and quasars (right), 
$\star$ BLRGs, $\downarrow$ upper limits, $\rightarrow$ 
lower limits. Statistics for the fits are listed in Table~\ref{tb-ASURV}.}
\label{lo-Rc-FRI-samp}
\end{figure}

For the FRI matched subsample, we restrict the redshifts to $z < 0.3$
and the extended radio luminosity at 1.4~GHz to $23.5\leq$~log$L_{ext}\leq25$ 
W Hz$^{-1}$. For the FRII matched subsample, the redshifts are 
constrained to $z < 1.3$, while the extended radio luminosity is 
$26.2\leq$~log$L_{ext}\leq27.6$ W Hz$^{-1}$. 
Figure~\ref{lo-Rc-FRI-samp} shows the $L_{o}$ -- $R_c$  
correlations for these subsamples while the correlation and regression 
parameters are listed in Table~\ref{tb-ASURV}.
We find that the scatter seen in Fig.~\ref{lo-rc-FRBQ} is considerably
reduced in Fig. 4 and the correlations improve significantly compared 
to the unrestricted samples.
Multiple linear regression tests on the restricted 
samples with the independent variables, $R_c$, $z$, $L_{ext}$
show that the $L_{o}$ -- $R_c$ is still the strongest correlation 
($p < 0.0001$) while the contribution of $L_{ext}$ is no longer significant.

\begin{table}[h]
\begin{center}
\caption{Statistics of correlations.}
\begin{tabular}{llccccc}\hline\hline
Type          &N~($l_x,l_y$)&X   &Y       &Spearman&Kendall&Schmitt(slope, intercept)\\
\hline
FRI galaxies  &54~(5,17) &log $R_c$&log $L_o$&0.0001 &0.0001 &0.52(0.17)~~19.40(0.25)\\
BL~Lacs       &44~(6,0 ) &log $R_c$&log $L_o$&0.0295 &0.0323 &0.17(0.14)~~22.45(0.16)\\
FRI \& BL~Lacs&98~(11,17)&log $R_c$&log $L_o$&$<$0.0001&$<$0.0001&1.15(0.10)~~20.72(0.14)\\
FRI-BL sample&57~(1,2)&log $R_c$&log $L_o$&$<$0.0001&$<$0.0001&1.20(0.17)~~20.72(0.16)\\
FRII galaxies &42~(1,20) &log $R_c$&log $L_o$&0.2175 &0.2392 &0.32(0.23)~~20.12(0.38)\\
BLRGs \& QSRs &47~(5,0)  &log $R_c$&log $L_o$&$<$0.0001&$<$0.0001&0.60(0.12)~~23.15(0.12)\\
FRII \& QSRs  &89~(6,20) &log $R_c$&log $L_o$&$<$0.0001&$<$0.0001&1.18(0.11)~~22.26(0.17)\\
FRII-QS sample&38~(0,3)&log $R_c$&log $L_o$&$<$0.0001&$<$0.0001&1.12(0.11)~~22.71(0.15)\\
\hline
FRI galaxies$^V$&9~(0,2) &$b/a$~$(d)$  &log $L_o$&0.2499$^a$&0.2310&1.83(1.22)~~17.67(0.89)\\
~~~~~~~~"      &14~(0,5) &$b/a$~$(d+l)$&log $L_o$&0.0928$^a$&0.0399&1.91(0.61)~~17.54(0.37)\\
~~~~~~~~"      &9~(0,0)  &$b/a$~$(d)$  &log $R_c$&0.0732$^a$&0.0953&--1.09(0.47)~~--0.19(0.37)\\
~~~~~~~~"      &14~(0,0) &$b/a$~$(d+l)$&log $R_c$&0.6065$^a$&0.5200&--0.37(0.40)~~--0.80(0.23)\\
FRI galaxies$^D$&6~(0,0) &$b/a$~$(d)$  &log $L_o$&0.0845$^a$&0.0909&2.24(0.82)~~17.62(0.48)\\
~~~~~~~~"      &7~(0,0)  &$b/a$~$(d+l)$&log $L_o$&0.0543$^a$&0.0509&2.33(0.76)~~17.63(0.54)\\
~~~~~~~~"      &7$^\star$~(0,0)  &$b/a$~$(d)$  &log $R_c$&0.0802$^a$&0.0985&1.39(0.53)~~--2.30(0.29)\\
~~~~~~~~"      &8$^\star$~(0,0)  &$b/a$~$(d+l)$&log $R_c$&0.0588$^a$&0.0833&1.51(0.48)~~--2.30(0.29)\\
FRI galaxies$^{V+D}$&12~(0,2) &$b/a$~$(d)$&log $L_o$&0.1855$^a$&0.1531&1.53(0.70)~~17.95(0.44)\\
~~~~~~~~"      &18~(0,5) &$b/a$~$(d+l)$&log $L_o$&0.0353$^a$&0.0248&1.90(0.60)~~17.66(0.33)\\
~~~~~~~~"      &12~(0,0) &$b/a$~$(d)$  &log $R_c$&0.7630$^a$&0.6808&--0.06(0.69)~~--1.06(0.49)\\
~~~~~~~~"      &18~(0,0) &$b/a$~$(d+l)$&log $R_c$&0.9085$^a$&0.9698&0.02(0.48)~~--1.12(0.28)\\
\hline    
\end{tabular}
\label{tb-ASURV}
\end{center}
Statistical significance of various correlations (of X and Y) and linear
regression fits. All the results are derived using ASURV as implemented 
in IRAF. Col.~1: the subclass of objects under consideration, `FRII 
galaxies' refer to narrow-line FRIIs alone, `QSRs' refer to quasars, 
`FRI-BL' and `FRII-QS samples' refer to the matched subsamples 
of FRIs and FRIIs as described in Sect.~\ref{secmatched}.
FRI galaxies with superscripts $V$, $D$, $V+D$ refer to FRI sources 
from \citet{Verdoes99}, \citet{deKoff00} and from both papers, 
respectively; $\star$ an additional FRI source -- 3C~430 with a disk of 
$b/a$ = 0.15 and log$R_c$ = -- 2.5 was included in the 
$b/a$ -- log$R_c$ correlation; 
Col.~2: the number of data points and those with limits in X and 
Y respectively, in paranthesis;
Col.~3 \& 4: the independent and dependent variable respectively; 
$b/a$ being the ratio of the minor-to-major axis of the extended 
dust feature seen in the {\it HST} images of radio galaxies, 
`$d$' and `$l$' standing for a dust disk and a lane respectively, 
`$d+l$' refers to our jointly considering disks and lanes in the 
correlations; 
Col.~5 \& 6: probability that no correlation exists between X and Y from 
Spearman's $\rho$ and Kendall's $\tau$ correlation tests respectively; 
`a' - Spearman Rank test is not accurate as no. of objects, N $<$ 30; 
Col.~7: slope and intercept with standard deviation in parantheses
from Schmitt's linear regression test, bootstrap approximation using 200
iterations, X bins = 10, Y bins = 10.
\end{table}

\subsection{Model-fitting the $L_{o}$ -- $R_c$ data}

If bulk relativistic motion with a single Lorentz factor ($\gamma$) value 
applicable to the whole population were alone responsible for the 
variation in $L_{o}$, then the logarithmic plot of $L_{o}$ against $R_c$ 
would be linear. Any additional factors such as orientation effects due to a 
torus or thin thermal disk will cause this relationship to deviate from 
linearity. We attempt to fit some simple models to the data along these 
lines. Appendix~\ref{appmodel} gives the model equations while 
Appendix~\ref{appfit} describes the different models considered along with 
the model-fitting procedure. 

We assume that the nuclear optical luminosity $L_{o}$ is, in the most
general case, due to the sum of synchrotron emission from the base of
a relativistic jet, and thermal emission from a thin accretion disk,
modified by the presence of an optically thick torus. Keeping in mind that 
we are only attempting to explore the potential of such a model-fitting 
approach and that our sample is not rigorously selected, we consider this 
simple model here and do not include the possibility of variation in 
{\it intrinsic} nuclear power as discussed in Sect.~\ref{secpop}, nor the possibility 
of extinction of the optical nucleus by an extended kpc-scale dusty disk 
which was discussed in Sect.~\ref{secdust}. 
Our models also do not take into account any intrinsic spread in the 
Lorentz factors, nor the possibility that the relevant Lorentz factor for 
the highly beamed and mildly beamed subclasses may be systematically 
different due to a ``spine-sheath" type structure of the jet 
\citep[e.g.,][]{Hardcastle96,Laing99}. 
However, as the multiple linear regression tests discussed in
Sect.~\ref{secpop} suggest, orientation appears to play the most dominant role in 
the variation of $L_{o}$. 
To quantify the goodness-of-fit of a particular model, we 
have used Akaike's information criterion (AIC). 
The AIC \citep{Akaike74} is a likelihood criterion with an added 
penalty term corresponding to the complexity of the model, and measures 
the trade-off between model complexity/parsimony and goodness-of-fit.
Smaller AIC values indicate a better fit.
Appendix~\ref{appfit} describes the usage
of AIC to derive the best model-fits to the FRI and FRII data.

\begin{table}[h]
\caption{Parameters from the different model-fits for the FR populations.}
\begin{center}
\begin{tabular}{l|lcc|ccccc}\hline\hline
&      &         &             &               &~~~~Model&Outputs &   &    \\
& Model&$A_{V_0}$&$\theta_c$   &$L_{jet}^{int}$&$\sigma$&$L_{disk}$&$\sigma$& AIC\\\hline
&Jet only & ...     & ...      &2.7e+20   &1.3e+19&  ...    &   ... & 295.2\\
&Jet+Disk & ...     & ...      &2.6e+20   &1.4e+19& 8.5e+17 &1.1e+18& 297.1\\
FRI&Jet+Disk+Torus&3.0&45      &2.7e+20   &1.5e+19& 3.5e+18 &1.6e+18& 304.5\\
&Jet+Torus& 3.0     & 45       &2.9e+20   &1.4e+19&   ...   & ...   & 303.1\\
&Best fit & 0.1     & 90       &2.6e+20   &1.3e+19&   ...   &  ...  & 295.3\\\hline
&Jet only          &...&...    &3.8e+20   &2.5e+19&   ...   & ...   & 132.3\\
Matched FRI&Jet+Disk&...&...&4.1e+20&2.9e+19&--7.9e+18 &2.0e+18& 133.4\\
&Jet+Disk+Torus    &3.0&45     &4.4e+20   &3.3e+19&--1.4e+18 &4.1e+18& 137.0\\
&Jet+Torus         &3.0&45     &4.4e+20   &2.9e+19&   ...   & ...   & 135.0\\\hline

&Jet only &  ...    &  ...     &5.8e+21   &3.0e+20&   ...   & ...   & 269.6\\
&Jet+Disk &  ...    &  ...     &6.5e+21   &3.7e+20&--6.7e+19 &1.2e+19& 270.7\\
FRII&Jet+Disk+Torus&3.0&45     &6.7e+21   &3.9e+20&--5.8e+19 &1.5e+19& 272.9\\
&Jet+Torus& 3.0     &  45      &6.1e+21   &3.2e+20&   ...   &  ...  & 271.5\\
&Best fit & 3.0     &  37      &6.5e+21   &3.4e+20&   ...   &  ...  & 270.8\\\hline

&Jet only          &...&...    &1.6e+21   &1.3e+20&   ...   & ...   & 94.3\\
Matched FRII&Jet+Disk&...&...&1.8e+21&1.5e+20&--1.9e+20&2.1e+19&93.4\\
&Jet+Disk+Torus    &3.0&45     &2.0e+21   &1.7e+20&--2.1e+20 &2.5e+19& 95.9\\
&Jet+Torus         &3.0&45     &1.8e+21   &1.4e+20&   ...   & ...   & 96.2\\\hline

BL FRII   &Jet only & ...&...  &2.2e+22   &1.6e+21&   ...   &  ...  & 127.3\\
          &Jet+Disk & ...&...  &1.3e+22   &1.2e+21& 1.5e+22 &2.7e+21& 123.3\\\hline
BL~Lacs$^\star$&Jet only&...&...&1.2e+21  &9.0e+19&   ...   &  ...  & 118.4\\
          &Jet+Disk & ...&...  &6.2e+19   &1.4e+19& 3.3e+22 &3.0e+21& 78.6 \\\hline
\end{tabular}
\label{tb-LMFIT}
\end{center}
$\star$ See Sect.~\ref{secresults} in the text.
In Col.~1 FRI/FRII and Matched FRI/FRII stand for the 
FRI/FRII population and it's matched subsample as discussed in 
Sect.~\ref{secmatched}, 
BL FRII stand for the broad-line FRIIs {\it viz.,} BLRGs and quasars. 
$A_{V_0}$ and $\theta_c$ 
(in degrees) are the fixed initial parameters for the models where a torus 
is incorporated. $\sigma$ is the standard deviation for the variable on 
the left. A lower AIC (Akaike's Information Criterion) value indicates 
a better model fit. The `Best fit' model is the `Jet+Torus' model for FRIs 
and FRIIs for which AIC is lowest (see Appendix~\ref{appfit} and
Sect.~\ref{secresults}), the 
$A_{V_0}$ and $\theta_c$ (in degrees) are the parameters corresponding 
to this fit.
\end{table}

\subsubsection{The results}
\label{secresults}

The results of the model-fitting are given in Table~\ref{tb-LMFIT}. 
For the FRI population, the `Jet-only' model is better than all the others. 
The `Jet+Disk' model in fact yields a value for the accretion disk
luminosity $L_{disk}$, which is 
comparable to its standard deviation $\sigma$ obtained from the fitting. 
Further, the `Jet+Torus' model is best fitted by the torus extinction
parameter $A_{V_0}$ $\approx$ 0.1 and half-opening angle 
$\theta_c$ $\approx$ 90$\degr$, which are equivalent to 
there being no torus. We note here that, based on the high rate of  
detections of optical nuclei in FRIs, \citet{Chiaberge99} have also suggested
that there is no obscuring torus in them. Further, \citet{Perlman01} have 
failed to detect thermal emission from a dusty torus
in the 10 $\mu$m image of the nearby FRI radio galaxy, M87.

\begin{figure}[h]
\centerline{
\includegraphics[height=6.5cm]{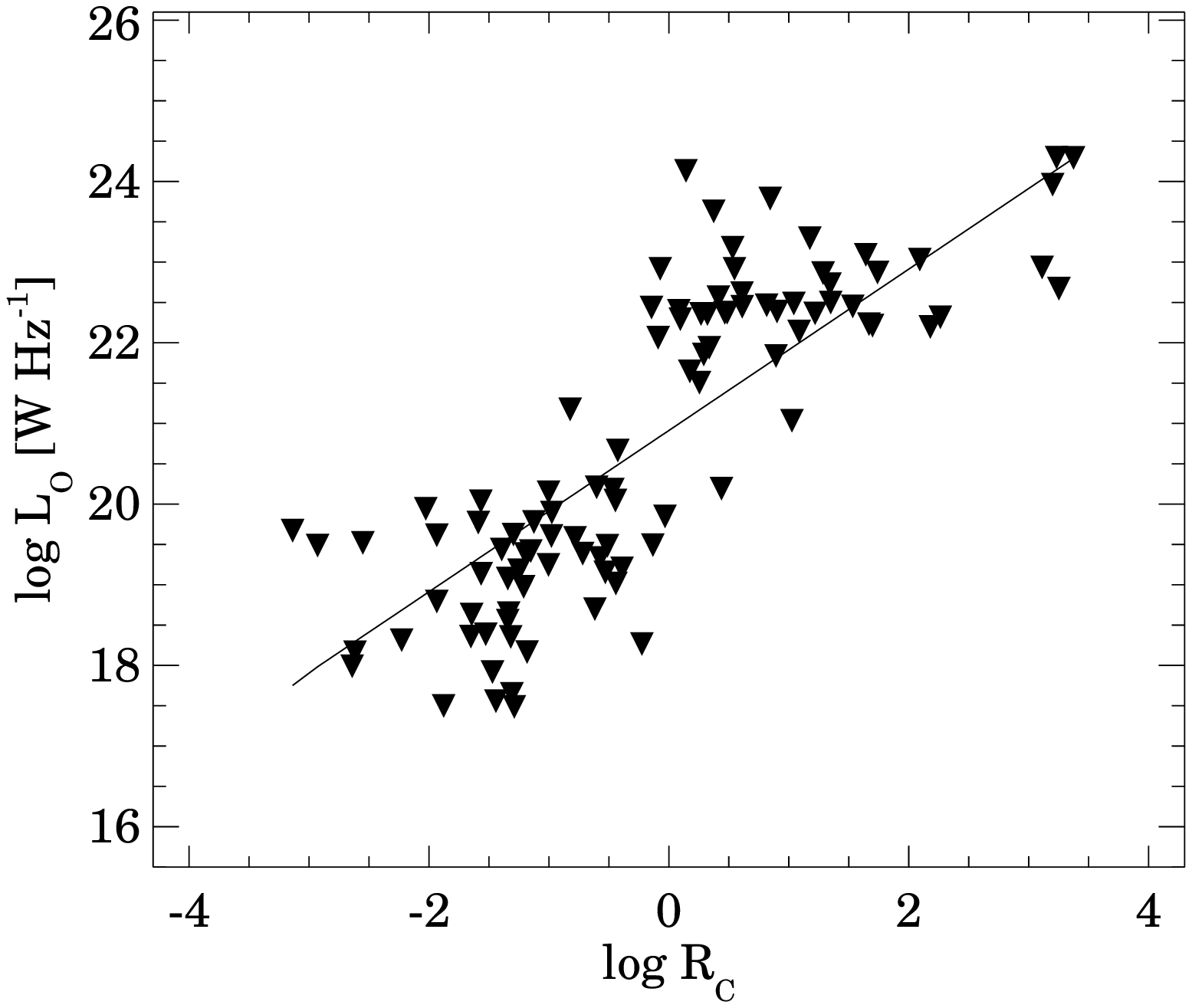}
\includegraphics[height=6.5cm]{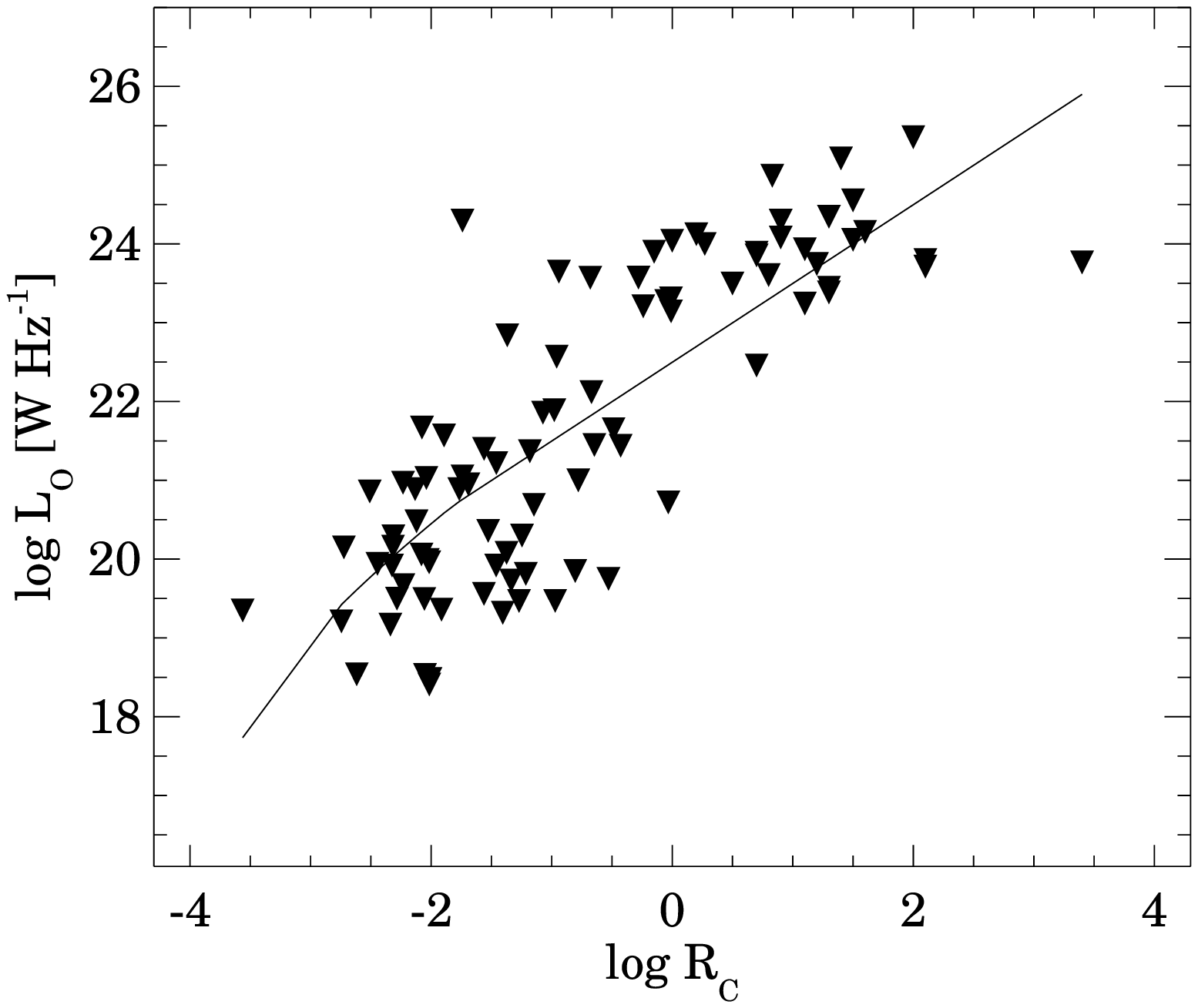}}
\caption{Best fits to the FRI (left) and FRII (right) populations using only 
a `Jet+Torus' model. Table~\ref{tb-LMFIT} lists the model parameters.}
\label{best-fit}
\end{figure}

For the FRII population as a whole, the results are less clear. Formally, 
the `Jet-only' model has the lowest AIC value, but the other 
models also yield comparable values. 
However, the $L_{disk}$ that is obtained for `Jet+Disk' 
and `Jet+Disk+Torus' is unphysical. 
When only the broad-line objects among the FRIIs, {\it viz.,} the BLRGs 
and quasars are considered, the `Jet+Disk' model was a better fit than 
the `Jet-only' model (Table~\ref{tb-LMFIT}). This is consistent with the 
fact that the `big blue bump' (attributed to the accretion disk) is 
observed in all these objects, and, in the framework of the US, 
their central regions are not obscured by the torus.
For the whole population, a larger number of data points in the regime 
where the disk is expected to be relatively most prominent, $viz.,$ the 
intermediate $R_c$ region, is required to derive a more 
robust quantitative 
value for $L_{disk}$ since at large $R_c$ the jet overwhelms the disk 
emission and at very small $R_c$ the torus obscures it.
Table~\ref{tb-LMFIT} lists the fitted parameters 
of the `Best fit' model for both the FR populations and Fig.~\ref{best-fit} 
shows the best fit curves to each of the FR populations. 

\begin{figure}[h]
\centerline{
\includegraphics[height=6.5cm]{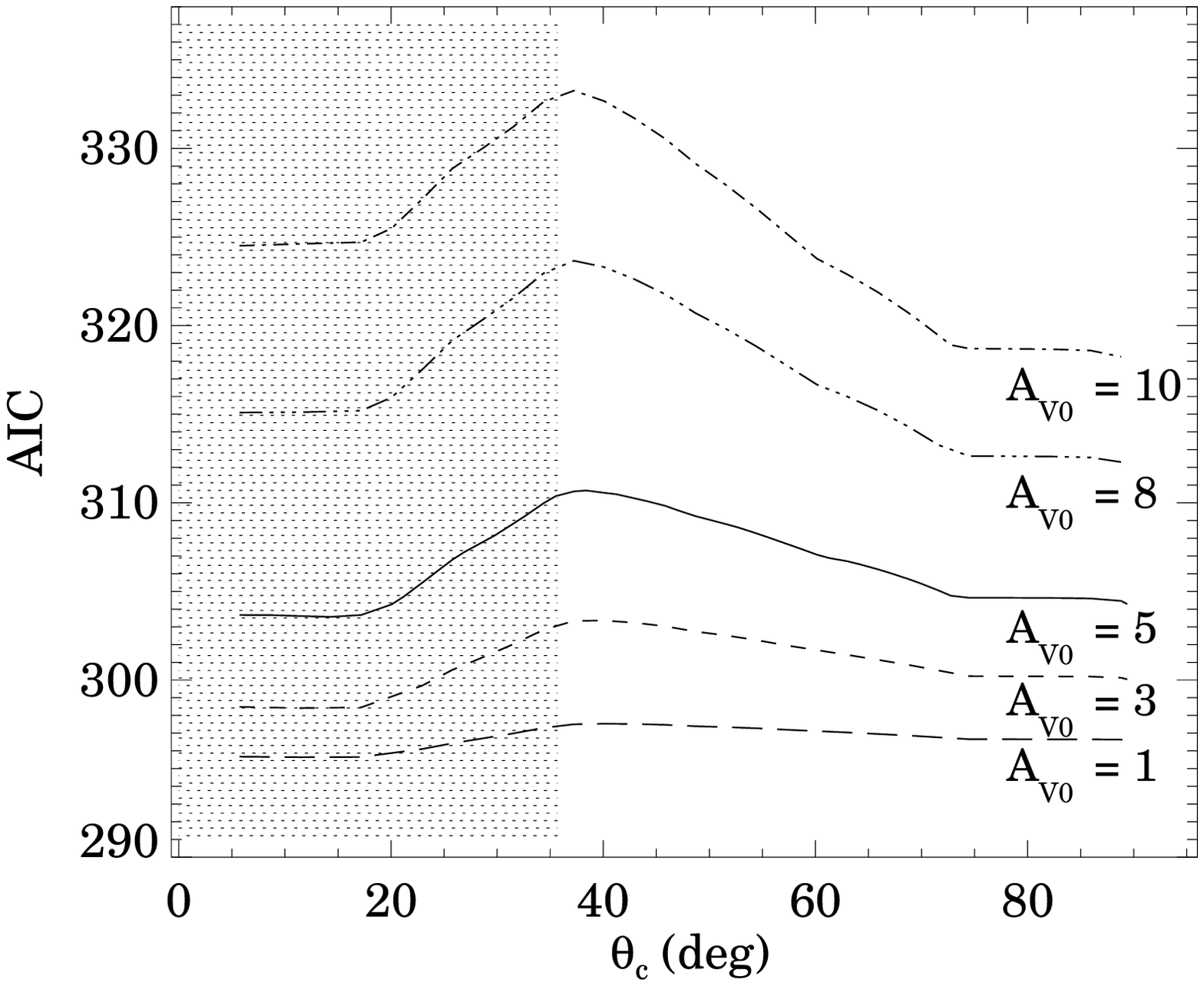}
\includegraphics[height=6.5cm]{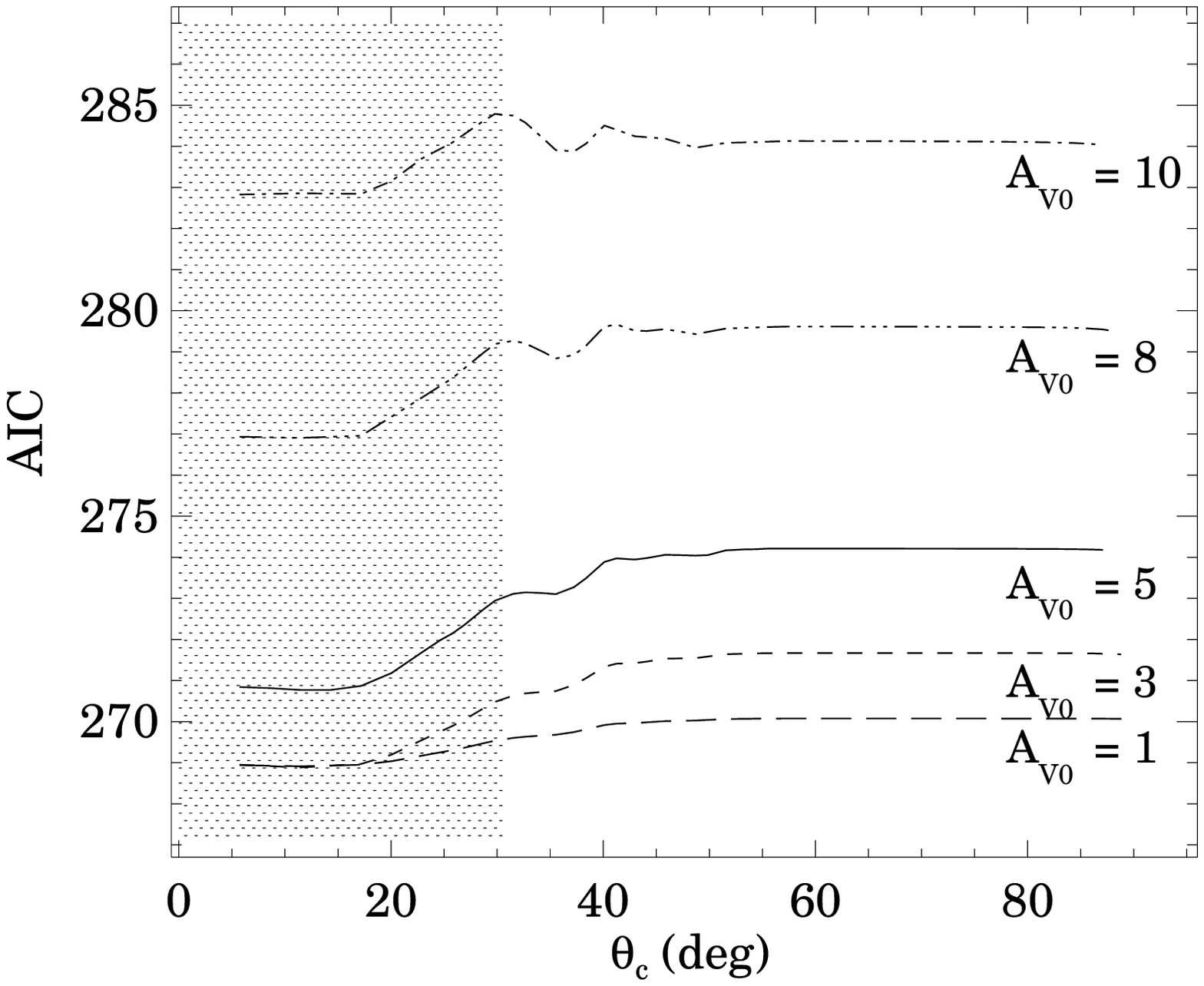}}
\caption{AIC values for different initial $A_{V_0}$ plotted against 
torus opening angles $\theta_c$ (in degrees) for the FRI (left) and 
the FRII (right) populations for the `Jet+Torus' model.
For a given $A_{V_0}$, AIC was estimated at 2$\degr$ intervals of
$\theta_c$.  
The shaded area denotes the region where the model becomes unphysical
(see Sect.~\ref{secresults}). In the physical regime, AIC 
reaches a minimum 
at around 90$\degr$ for the FRIs and 37$\degr$ for the FRIIs.}
\label{AIC}
\end{figure}

It is interesting here, that the `Jet-only' model is {\it unambiguously 
the best-fit for the FRIs}, whereas, {\it several models give 
comparable fits to the FRIIs.} Given this difference in the 
behaviour of the two classes,
we carry the procedure a bit further by contrasting the behaviour of 
the AIC for the FRIs and FRIIs in the `Jet+Torus' model case.
In Fig.~\ref{AIC} we plot the AIC against the torus half-opening angle for 
different fixed values of $A_{V_0}$. Figure~\ref{AIC} shows that the 
families of 
AIC plots for the two populations differ systematically from each other. 
The plots can be broadly divided into two parts. Below 
$\theta_c \approx$ 30$\degr$ 
the AIC drops for both the FRI and FRII population (shaded region in
Fig.~\ref{AIC}). This formally implies 
that the fit gets better for opening angles of the torus that are smaller 
than $\theta_c \approx$ 30$\degr$, but clearly is the result of the 
algorithm 
trying to fit the {\it entire} variation in $L_o$ by torus obscuration 
{\it alone}. For the FRIs,
above $\theta_c \approx$ 35$\degr$, the AIC declines again and reaches 
a minimum at angles close to 90$\degr$, 
consistent with there being no torus. For the FRIIs, on the other hand
the AIC does not decline appreciably 
above $\theta_c \approx$ 35$\degr$. In addition it shows a 
conspicuous minimum at $\theta_c \approx$ 37$\degr$. 

Thus although simple model-fitting using the LM algorithm yields 
ambiguous results for the FRIIs on the face of it, by rejecting the 
possibility of the entire variation in $L_o$ being due to obscuration 
by a torus we obtain model parameters that are broadly consistent with 
the predictions of the US. The model-fitting results in the best 
$\theta_c$ roughly coinciding with the angle where the upper limits 
to $L_o$ start appearing in FRII galaxies. 
The best $A_{V_0}$ turns out to $\approx$ 3~mag for the FRIIs at the 
best $\theta_c$ while for the FRIs, $A_{V_0}$ turns out to be $\approx$~0. 
We point out that the $A_{V_0}$ that we infer here is of the nature of 
a lower limit, since detection limits would exclude measured data points 
corresponding to higher values of $A_{V_0}$.
We find that the behaviour of FRI and FRII optical nuclei 
is distinctly different in that the model-fitting results are unambiguous 
for the FRIs while they are not so for the FRIIs, hinting at intrinsic 
differences between FRIs and FRIIs.

It may be recalled that the BL~Lac objects taken by themselves show a flatter 
logarithmic distribution of $L_o$ against $R_c$ than would be expected from 
beaming alone (Sect.~\ref{secpop}), 
and indeed model fitting just the BL~Lacs gives the 
`Jet+Disk' model to be the best one for them, with an implied $L_{disk}$ of 
$3.3\times10^{22}$ W Hz$^{-1}$ (See Table~\ref{tb-LMFIT}). As was stated in 
Sect.~\ref{secpop}, this is most likely to be due to the contaminating host galaxy 
luminosity mimicking emission from a disk. We note however, that unless 
values for their optical nuclei are used that are uncontaminated 
by the host galaxy, 
we cannot totally rule out an accretion disk in the BL~Lacs, 
and this implies that they may not be intrinsically similar to 
the FRI radio galaxies.

\begin{figure}[h]
\centerline{
\includegraphics[height=6.5cm]{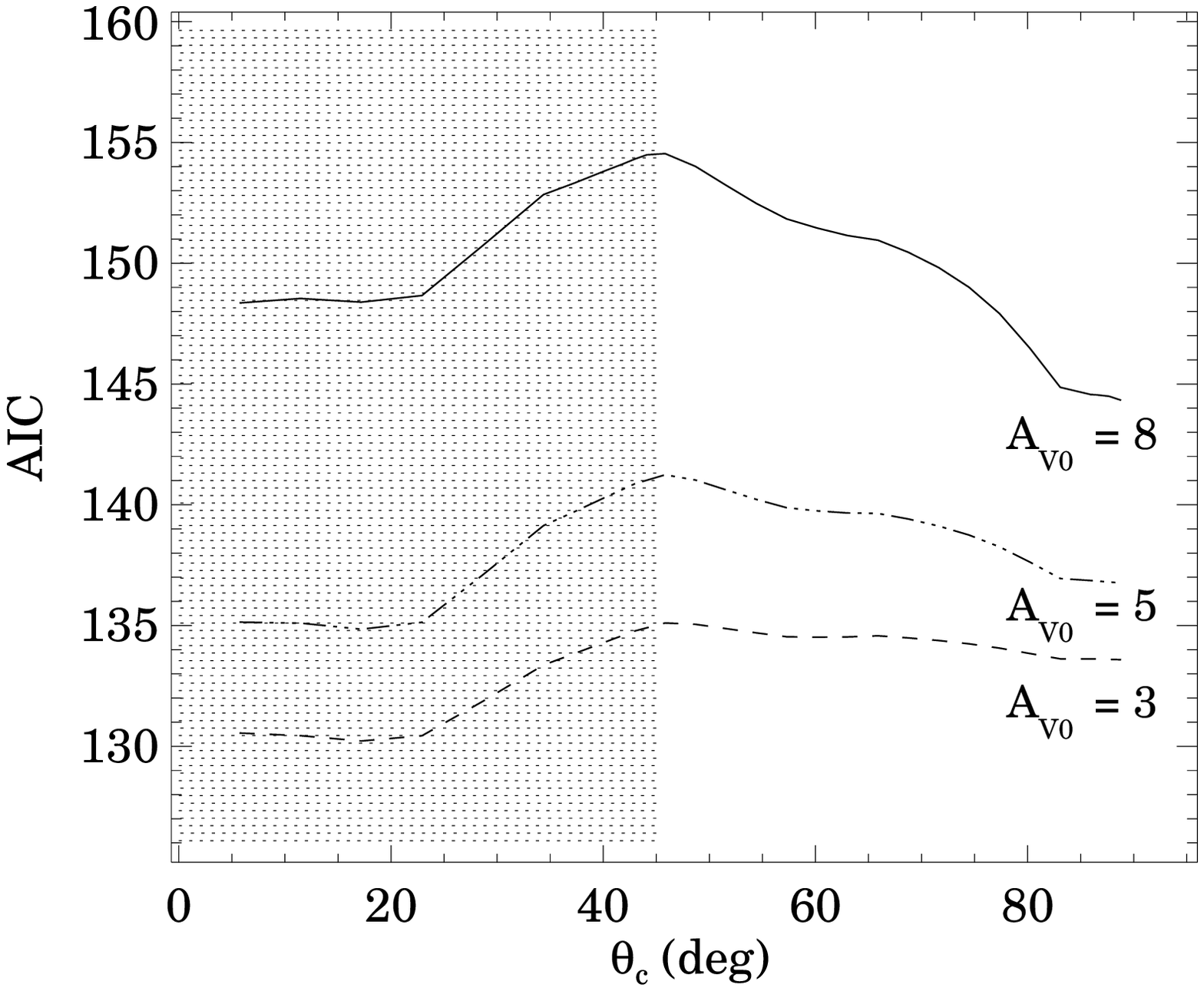}
\includegraphics[height=6.5cm]{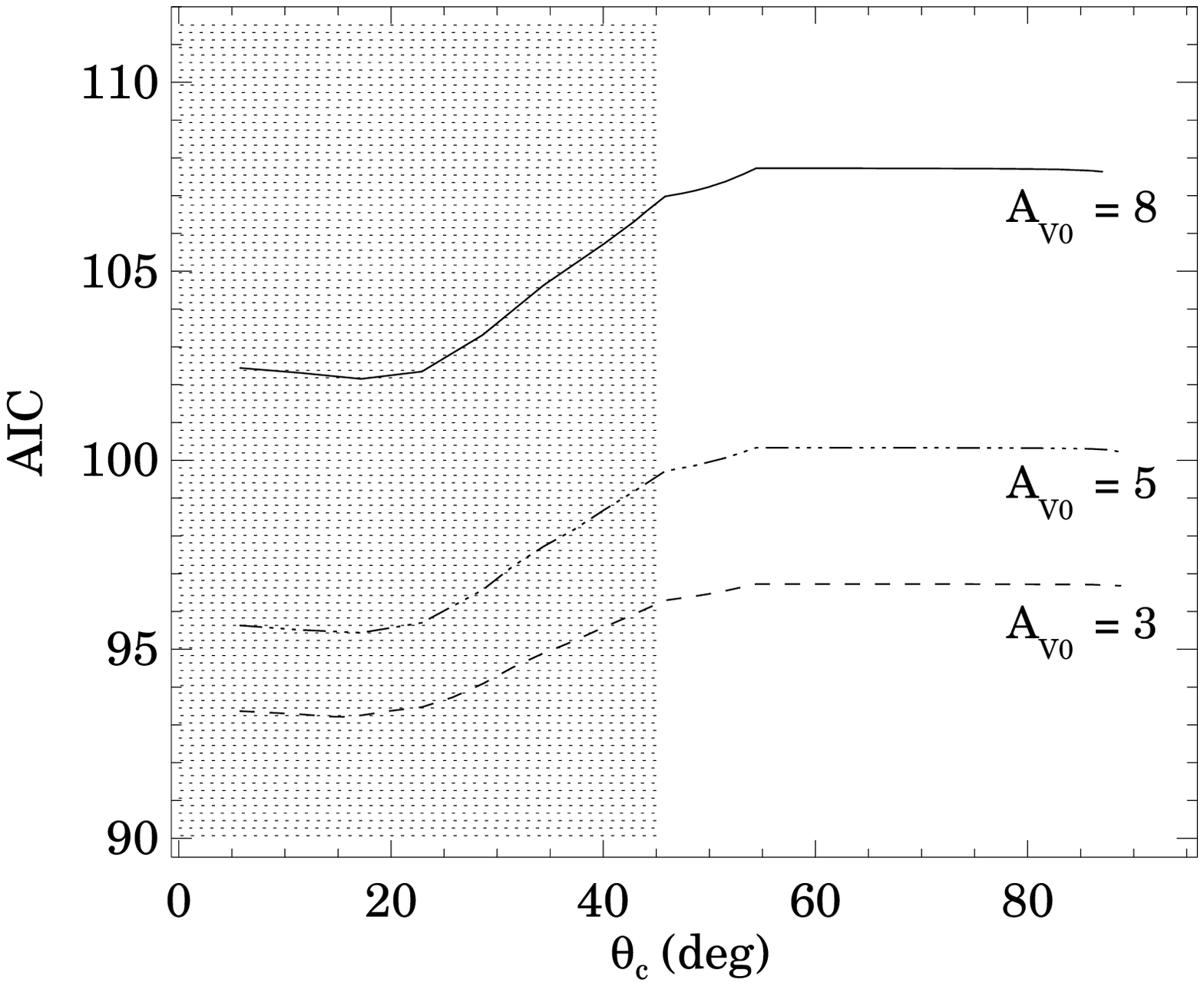}}
\caption{AIC values for different initial $A_{V_0}$ plotted against
torus opening angles $\theta_c$ (in degrees) for the matched subsamples 
of FRI (left) and the FRII (right) populations for the `Jet+Torus' model.
The shaded area denotes the region where the model becomes unphysical
(see Sect.~\ref{secresults}). In the physical regime, AIC
reaches a minimum
at around 90$\degr$ for the FRIs while the case is not clear for
the FRIIs.}
\label{AIC-sampl}
\end{figure}

For the $smaller$ matched subsamples of FRIs and FRIIs, 
the variation of AIC with $\theta_c$ (see Fig.~\ref{AIC-sampl}) is 
broadly similar to that of their respective unrestricted samples. 
This validates our procedure and the {\it qualitative} results for
the whole sample regarding the differences between the FRIs and FRIIs. 
The only discernible difference for the `matched' FRIIs is that 
the `AIC minimum' observed at $\theta_c$ $\approx 37\degr$ for the 
whole sample, is now no longer prominent. However, this may not be
surprising in view of the fact that the `matched' FRII objects are much fewer
in number, especially at low $R_{c}$ values.

We note that for neither the FRIs nor FRIIs 
is there any correlation between the residuals of the model-fit
with $R_{c}$ and $L_o$. However, there seems to be a weak correlation for 
the BL~Lacs considered alone. This may reflect the effects of 
ignoring the host galaxy luminosity when taking the nuclear optical 
luminosity.  

We note that while reliable quantitative results cannot be 
obtained using the current data because of the drawbacks in the sample, 
the approach and the results indicate that it is a good approach to derive 
various parameters if data for a rigorous and large sample are available. 
Better data would allow more parameters to be incorporated and controlled.

\subsubsection{The bulk Lorentz factor}

As an approximation we assume that a single $\gamma$ value is applicable 
to each population, and that all orientations are represented in 
each population, and therefore that the minimum and maximum values of 
$R_c$ in each population correspond to orientations perpendicular 
and parallel to our line of sight respectively.
The formula using the $R^{min}_c$ and $R^{max}_c$ (Eq.~\ref{eq3} in 
the Appendix)
results in the lower limit to the maximum Lorentz factor $\gamma_{max} 
\approx$ 9.7 for the FRI population. For the FRII population we get 
a value of $\gamma_{max} \gtrsim 11.5$, obtained using a quasar with a 
upper limit to its extended radio emission, and therefore the actual lower 
limit to $\gamma_{max}$ could be higher. These values broadly agree with 
those obtained by \citet{UrryPadovani95} : $\gamma_{max} \gtrsim 9$ for 
the FRI population and $\gtrsim 13$ for the FRII population assuming $p = 3$. 
However, based on the correlation of optical and radio core emission with 
the isotropic H$\alpha$+[NII] emission, \citet{Verdoes02} have obtained a 
constraint on the value of $\gamma$ of $\lesssim$~2 (assuming $p~=~3$), 
albeit for FRI radio galaxies {\it alone}.

\section{Conclusions}

We use the radio core prominence $R_{c}$ as a statistical indicator 
of orientation and find that the systematic differences between radio-loud 
AGN of the two Fanaroff-Riley types appear to also extend to their optical 
nuclei, in a manner that is consistent with the predictions of the simple US.
We find that\\
1.~~The luminosity of the pc-scale optical nuclei in the FRI radio 
galaxies is orientation-dependent, while that in the FRII radio galaxies 
is not. This result is consistent with the idea that FRIIs contain an 
obscuring torus, (as required by the simple US) whereas there is no torus 
in the FRIs.\\
2.~~For the FRI radio galaxies, though the correlation with orientation is 
very significant, there remains considerable residual scatter. 
This may be due to obscuration from an 
{\it extended kpc-scale} dusty disk. The axis of 
this disk appears unrelated to the AGN axis. The residual scatter may also 
be due to intrinsic variation in the optical luminosity.\\
3.~~The nuclear optical luminosity correlates significantly with $R_{c}$, 
or, equivalently orientation, for the FRI radio galaxies and BL~Lacs of FRI 
morphology taken together -- the FRI population. Our model-fitting suggests 
that a relativistically beamed optical jet gives the best fit.\\ 
4.~~For the FRII radio galaxies and radio-loud quasars taken together 
-- the FRII population, the nuclear optical luminosity again correlates 
significantly with $R_{c}$. Our model-fitting indicates that formally the 
best fit is again a beamed synchrotron jet. But a beamed jet obscured by a 
torus with an inferred opening angle close to 40$\degr$ is a comparable fit, 
and is able to explain the contrasting behaviour of the FRI and FRII data.\\
5.~~The scatter in the $L_o$ -- $R_c$ correlation for both the
FR populations is likely to be primarily due to the spread in 
intrinsic AGN power, although extended dusty disks may also contribute
to the scatter in FRI radio galaxies.\\
6.~~Our model-fitting suggests that the luminosity of the intrinsic 
({\it i.e.,} unbeamed) jet in the FRIIs is approximately an order of magnitude 
larger than for the FRIs, although this result needs to be confirmed 
using a rigorous sample.\\
7.~~The data for the broad-line FRIIs alone are fitted best 
by a model that 
comprises a relativistic jet and a geometrically
thin optically thick disk, consistent 
with the presence of the `big blue bump' in them.

The robustness of the above results is 
limited by the facts that (a) the 
``samples" used are eclectic, (b) the 
luminosities of the optical nuclei in the 
highly beamed objects are contaminated by the contributions 
of the host galaxies, 
and (c) there could be variability between the epochs of the 
optical and radio 
measurements. A robust analysis requires rigorous measurement of the optical 
luminosity as well as samples that are rigorously selected, with the objects of 
a given FR population chosen to be intrinsically similar in the framework of 
the US and from the same volume of space. 

\begin{acknowledgements}
We are very grateful to Prof. Thriyambakam Krishnan for his 
scrutiny of our statistical tests and results, useful suggestions 
and extensive discussions. We also thank the anonymous referee for 
comments that improved the paper.
\end{acknowledgements}

\appendix
\section{The model equations}
\label{appmodel}

We give here the equations that form the basis of our model-fitting procedure.
In the most general case, we assume that the nuclear optical 
luminosity $L_{o}$ is due to the sum of synchrotron emission from the base of 
a relativistic jet, and thermal emission from a 
thin accretion disk, 
modified by the presence of an optically thick torus. 
We write, 
\begin{equation}
L_o = (\delta^{p}~L_{jet}^{int} + 
L_{disk}~cos~\theta)\times10^{-A_V/2.5}
\label{eq1}
\end{equation}

$L_{jet}^{int}$ is the intrinsic synchrotron luminosity from the base 
of the jet which is relativistically beamed by the factor 
$\delta^p$, where $\delta$ is the Doppler factor and for a jet spectral 
index of $\alpha$ the jet structure parameter $p$ 
is given by 2+$\alpha$ or 3+$\alpha$ depending on whether the jet is continuous 
or blobby \citep[e.g.,][]{UrryPadovani95}. 
$L_{disk}$ is the luminosity of a geometrically thin 
optically thick accretion disk, 
whose apparent luminosity is orientation-dependent due to projection 
(the  $cos~\theta$ term). $A_V$ is the extinction resulting from the 
torus in the {\it V} band. For a half-opening angle of the torus
$\theta_c$, we have,
 
\begin{equation}
A_V = A_{V_0}~\left(1 - \frac{cos~\theta}{ cos~\theta_c}\right) \hspace{2cm}
{\rm for} \hspace{5mm} \theta \ge \theta_c \hspace{1cm} 
\label{eq2}
\end{equation}
$A_V = 0$ \hspace{4.1cm} for \hspace{4mm} $\theta < \theta_c$ \\
\citep{Simpson96}\\
Thus, for $\theta = 90\degr$ $ A_V = A_{V_0}$. 

For the Lorentz factor of bulk relativistic motion of the nuclear jet 
\citep[e.g., Appendix C,][]{UrryPadovani95} we have, 

\begin{equation}
\gamma = \left(\frac{1}{2^{p-1}} 
\frac{R^{max}_c}{R^{min}_c}\right)^{\frac{1}{2p}},
\label{eq3}
\end{equation}
\begin{equation}
R_{c}^{int} = \frac{\gamma^p R^{min}_c}{2}
\label{eq5}
\end{equation}

where $R^{min}_c$ and $R^{max}_c$ are the minimum and maximum values 
of $R_c$, {\it i.e.,} the values of $R_c$ at edge-on ($\theta \sim 90 \degr$)
and pole-on ($\theta \sim 0 \degr$) inclinations of the AGN 
respectively, and $R_{c}^{int}$ is the 
intrinsic flux density ratio of the core and the extended radio emission. 
We now obtain the orientation to the line of sight, $\theta$, in terms 
of $R_c$. We use the relativistic beaming formulae 
which take into account contributions from both the approaching 
and receding jet \citep[e.g.,][]{UrryPadovani95}. 

\begin{equation}
R_c = R_c^{int} \left[\frac{1}{[\gamma(1 - \beta cos \theta)]^p} + 
\frac{1}{[\gamma(1 + \beta cos \theta)]^p}\right] 
\label{eq6}
\end{equation}

We assume a value of 3 for the jet structure factor $p$. We note that 
\citet{UrryPadovani95} infer a $p$ value of $\approx$ 3 based on the 
observations of superluminal motion within our own galaxy by 
\citet{Mirabel94}. We get,
 
\begin{equation}
\beta cos \theta = \sqrt{1 - 
\frac{(2 b)^{2/3}}{\sqrt{R_c} (-2\sqrt{R_c} + \sqrt{2 b + 4 R_c})^{1/3}} + 
\frac{(2 b)^{1/3}(-2\sqrt{R_c} + \sqrt{2 b + 4 R_c})^{1/3}}{\sqrt{R_c}}}
\label{eq7}
\end{equation}
where $b$ ={\Large $\frac{R_c^{int}}{\gamma^p}$.}

\section{The models and the fitting procedure}
\label{appfit}

We describe here the model-fitting procedure and the use of the Akaike's 
information criterion (AIC) to evaluate the models. 
We considered several simple models to compare the behaviour of the  
optical nuclei in the FRI and FRII populations. For the `Jet-only' model, 
the entire nuclear optical luminosity is ascribed to synchrotron emission 
from a jet which is relativistically beamed. For the `Jet+Disk' model, 
the nuclear optical luminosity is modelled as a combination of a beamed 
synchrotron jet and a thin optically thick disk. For the `Jet+Disk+Torus' 
model, Eq.~\ref{eq1} is used {\it in toto}. In the `Jet+Torus' model the entire 
nuclear optical luminosity is due to the beamed jet, modified by an 
obscuring torus. 

Using the $\gamma$ values derived as in Eq.~\ref{eq3} from the minimum 
and maximum values of $R_c$ in our data set for each of the FRI and FRII 
populations and Eqs.~\ref{eq1} and \ref{eq6} for $L_o$ and $R_c$, we made 
a non-linear least squares fit to the data for both the FR populations 
separately. We used the Levenberg-Marquardt (LM) algorithm as implemented 
in the IDL package (the LMFIT routine). This routine gives the best-fit 
values of the free parameters, their standard deviations and the $\chi^2$ 
goodness-of-fit. For the model-fitting we assumed the errors in the nuclear 
optical luminosity to be $50\%$ \citep[e.g.,][]{Verdoes02}.

The Akaike's information criterion is used to compare 
different model-fits and is defined as 
\begin{equation}
{\rm AIC = -2 ln(L) + 2 k \hspace{2cm}} 
\label{AICeq1}
\end{equation}
\citep{Burnham02}\\
where ln(L) is the log likelihood function and is given by 
\begin{equation}
{\rm ln(L) = -(n/2) \{ln(2 \pi) + ln(SEE/n) + 1\}}
\label{AICeq2}
\end{equation}
where `n' is the number of data-points, `SEE' is the Standard error of 
estimate and `k' is the number of parameters to be fitted. 
Smaller AIC values indicate a better fit. We note 
that the goodness-of-fit criterion AIC behaves in an inverse 
fashion to the $\chi^2$ probability $Q$ \citep{NumRec92}. 

For the `Jet-only' model, $L_{jet}^{int}$ is the only free parameter and the LM 
algorithm yields its best fit value. For the `Jet+Disk' model, both 
$L_{jet}^{int}$ and $L_{disk}$ are free parameters. The output of the algorithm 
turned out to be independent of the input seed values of these free 
parameters. The `Jet+Disk+Torus' and the `Jet+Torus' models have two 
additional parameters of the torus, {\it viz.,}, $A_{V_0}$ and $\theta_c$. 
However, apart from the quality of the data, the fact that there 
is more 
than one ``local minimum" for $\theta_c$ 
(discussed in Sect.~\ref{secresults}) 
did not allow 
a robust estimation of all the parameters simultaneously. (The estimated 
values for $\theta_c$ strongly depended on the input seed value). 
We therefore adopted the procedure of manually varying the extinction 
$A_{V_0}$ and torus half-opening angle $\theta_c$ and using the 
resulting AIC for each pair of (fixed) $A_{V_0}$ and $\theta_c$ to infer 
the best fit. 

Considering first the most general `Jet+Disk+Torus' model to the fit 
the data, we chose a range of values for the torus parameters, $A_{V_0}$ 
and $\theta_c$, and estimated the best fit $L_{jet}^{int}$ and $L_{disk}$ 
using the LM algorithm. 
The results are tabulated in Table~\ref{tb-LMFIT} for 
representative values, $A_{V_0}$ = 3 and 
$\theta_c$ = 45$\degr$ \citep[e.g.,][]{Barthel89}. 
The resulting value of $L_{disk}$ turned out to
be insignificant in each case (see Table~\ref{tb-LMFIT} and 
Sect.~\ref{secresults}). 
We therefore further considered only the `Jet+Torus' model which seemed more 
applicable to the data. For a given value of $A_{V_0}$ we varied the 
$\theta_c$ from 0$\degr$ through 90$\degr$ and tabulated the resultant AIC,
which are plotted in Fig.~\ref{AIC}.

As a next step, we fixed the torus half-opening angle $\theta_c$ to the 
value which had resulted in the minimum AIC (Fig.~\ref{AIC}) -- which is 
approximately the same for different values of $A_{V_0}$ for each FR 
population, and let the $A_{V_0}$ be the free parameter to be best-fitted 
by the LM algorithm. We found that the resultant $A_{V_0}$ for the FRII 
objects was independent of the seed value whereas it depended on the 
initial value for the FRI objects. In this manner we estimated the 
`Best fit' `Jet+Torus' model with values of $A_{V_0}$ and $\theta_c$ 
which gave the lowest AIC value. 

\section{Data on sample objects}
\begin{table}
\begin{center}
\caption{The FRI radio galaxies.}
\begin{tabular}{lcccccccccc}\hline\hline
IAU      &Alternate&Redshift&$b/a$&log$L_{ext}$&$S_{c}$(5~GHz)&ref.&log$R_c$&log$L_{o}$ &ref.\\
  name   &name     &  $z$   &     &W Hz$^{-1}$ &mJy           &    &        &W Hz$^{-1}$&  \\\hline
 0036+030& NGC 193  &0.0144 &0.18$^l$&22.60& 40.0   &6,15 &--1.31   & 18.35    &5\\
 0053+261& 3C 28    &0.1952 &...     &25.42&$<$0.2  &1,3  &$<$--2.92&$<$19.48  &1\\
 0055-016& 3C 29    &0.0448 &...     &25.29& 93.0   &1,3  &--1.39   & 19.43    &1\\
 0055+265& 4C 26.03 &0.0472 &...     &24.61& 9.0    &2    &--1.93   &$<$18.54  &2\\
 0055+300& NGC  315 &0.0167 &0.23$^d$&24.01& 617.6  &2    &--0.39   & 19.19    &5\\
 0104+321& 3C 31    &0.0169 &0.77$^d$&24.40& 92.0   &1,3  &--1.34   & 19.08    &5\\
 0120+329& NGC 507  &0.0164 &...     &22.29& 1.5    &2    &--1.30   &$<$17.64  &2\\
 0123-016& 3C 40    &0.0180 &0.91$^d$&22.17& 67.8   &10   &--1.64   &$<$18.63  &5\\
 0153+053& NGC 741  &0.0185 &...     &22.75& 6.0    &11   &--1.65   &$<$18.35  &5\\
 0220+427& 3C 66B   &0.0215 &0.98$^d$&24.86& 182.0  &1,3  &--1.29   & 19.62    &5\\
 0305+039& 3C 78    &0.0288 &...     &24.96& 964.0  &1,3  &--0.42   & 20.66    &1\\
 0318+415& 3C 83.1  &0.0251 &0.09$^d$&24.98& 21.0   &1,3  &--2.22   & 18.31    &1\\
 0316+413& 3C 84    &0.0176 &...     &24.73& 42370.0&1,3  & 1.02   & 21.03    &1\\
 0331-013& 3C 89    &0.1386 &...     &25.80& 49.0   &1,3  &--1.21   &$<$18.97  &1\\
 0705+486& NGC 2329 &0.0193 &0.68$^d$&23.02& 69.0   &11   &--0.44   & 20.04    &5\\
 0755+379& 3C 189   &0.0413 &...     &24.43& 228.8  &2    &--0.46   & 20.18    &2\\
 0924+301& ...      &0.0266 &...     &23.52&0.4  &2    &$<$--2.64&$<$17.98  &2\\
 0928+678& NGC 2892 &0.0225 &...     &22.82& 30.0   &12   &--0.52   & 19.15    &5\\
 1142+198& 3C 264   &0.0206 &0.99$^d$&24.57& 200.0  &1,3  &--1.00   & 20.15    &5\\
 1205+255& UGC 7115 &0.0226 &...     &22.58& 44.0   &11   &--0.51   & 19.48    &5\\
 1216+061& 3C 270   &0.0074 &0.46$^d$&24.31&308.0&1,3  &--1.44   & 17.56    &5\\
 1220+587& NGC 4335 &0.0154 &0.41$^d$&22.64& 15.0   &6    &--0.44   &$<$19.01  &5\\
 1222+131& 3C 272.1 &0.0037 &0.15$^l$&23.22& 180.0  &1,3  &--1.18   & 18.17    &5\\
 1228+126& 3C 274   &0.0037 &...     &24.63& 4000.0 &1,3  &--1.24   & 19.18    &5\\
 1257+282& NGC 4874 &0.0239 &...     &23.07& 1.2    &2    &--1.87   &$<$17.49  &2\\
 1322+366& NGC 5141 &0.0173 &0.25$^l$&23.63& 78.7   &10   &--0.61   &$<$18.69  &5\\
 1336+391& 3C 288   &0.2460 &...     &26.42& 30.0   &1,3  &--1.56   & 20.03    &1\\
 1346+268& 4C 26.42 &0.0633 &...     &24.52& 59.3   &2    &--0.78   & 19.58    &2\\
 1407+177& NGC 5490 &0.0162 &0.35$^l$&23.24& 37.8   &10   &--1.28   &$<$17.48  &5\\
 1414+110& 3C 296   &0.0237 &0.29$^d$&24.61& 77.0   &1,3  &--1.33   & 18.64    &1\\
 1422+268& ...      &0.0370 &...     &23.99& 21.1   &2    &--1.15   &$<$19.41  &2\\
 1430+251& 4C 25.46 &0.0813 &...     &24.20& 1.2    &2    &$<$--1.93&$<$19.61  &2 \\
 1450+281&    ...   &0.1265 &...     &24.48& 6.7    &2    &--1.12   &$<$19.77  &2\\
 1502+261& 3C 310   &0.0540 &...     &25.19& 80.0   &1,3  &--1.19   & 19.38    &1\\
 1510+709& 3C 314.1 &0.1197 &...     &25.33& $<$1.0 &1,3  &$<$--2.55&$<$19.52  &1\\
 1514+072& 3C 317   &0.0342 &...     &24.43& 391.0  &1,3  &--0.13   & 19.49    &1\\
 1521+288& 4C 28.39 &0.0825 &...     &24.53& 55.8   &2    &--0.60   & 20.20    &2\\
 1527+308&    ...   &0.1143 &...     &24.03& 4.0    &2    &--0.97   &$<$19.89  &2\\
 1553+245&    ...   &0.0426 &...     &23.43& 57.9   &2    &--0.03    & 19.84    &2\\
 1610+296& NGC 6086 &0.0313 &...     &22.92& 1.1    &2    &$<$--1.52 &$<$18.38  &2\\
 1613+275&    ...   &0.0647 &...     &24.01& 10.6   &2    &--1.00    &$<$19.24  &2\\
 1626+396& 3C 338   &0.0303 &...     &24.19& 105.0  &1,3  &--0.57    & 19.33    &1\\
 1637+826& NGC 6251 &0.024  &...     &23.82& 720.0  &13,4 &0.44     & 20.19    &7\\
 1641+173& 3C 346   &0.1620 &...     &26.20& 220.0  &1,3  &--0.82    & 21.17    &1\\
 1648+050& 3C 348   &0.1540 &...     &27.12& 10.0   &1,3  &--3.13    & 19.67    &1\\
 1827+323&    ...   &0.0659 &...     &24.04& 20.8   &2    &--0.71    &$<$19.38  &2\\
 2045+068& 3C 424   &0.1270 &...     &25.67& 18.0   &1,3  &--1.58    &$<$19.77  &1\\
 2116+262& NGC 7052 &0.0164 &0.30$^d$&22.97& 47.0   &9,14  &--0.22    & 18.26   &8\\
 2153+377& 3C 438   &0.2900 &...     &26.77& 17.0   &1,3   &--2.02    &$<$19.94 &1\\
 2212+135& 3C 442   &0.0262 &...     &24.39& 2.0    &1,3   &--2.61    & 18.16   &1\\
 2229+391& 3C 449   &0.0181 &0.50$^d$&24.29& 37.0   &1,3   &--1.56    & 19.13   &1\\
 2236+350& UGC 12127&0.0277 &...     &23.46& 7.1    &2     &--1.34    & 18.55   &2\\
 2318+079& NGC 7626 &0.0113 &0.17$^l$&22.39& 15.6   &10    &--1.47    &$<$17.91 &5\\
 2335+267& 3C 465   &0.0301 &0.69$^l$&25.00& 270.0  &1,3   &--0.97    & 19.60   &1\\
\hline
\end{tabular}
\label{tb-FRI}
\end{center}
Superscripts `$d$' and `$l$' for $b/a$ stand for extended dust 
disks and lanes, respectively.
References: (1) : \citet{Chiaberge99} (F702W filter); For the 7 sources 
which were common between the 3CR, B2 and the UGC samples we used the 
F555W flux densities from \citet{Verdoes02};
(2) : \citet{Capetti02} (1.4 GHz, F814W filter);
(3) : \citet{Kuehr79} (5 GHz);
(4) : \citet{Kuehr81} (5 GHz);
(5) : \citet{Verdoes02} (F555W filter);
(6) : \citet{Xu00} (1.4 GHz);
(7) : \citet{Hardcastle99} (F702W filter);
(8) : \citep{Capetti99} (F814W filter);
(9) : \citet{Giovannini88} (5 GHz);
(10) : \citet{BridlePerley84} (core at 5 GHz, total flux density at 1.4 GHz);
(11) : \citet{LaurentMuehleisen97} (5 GHz);
(12) : \citet{Jenkins82} (5 GHz);
(13) : \citet{Waggett77} (2.7 GHz);
(14) : \citet{Gregory91} (5 GHz);
(15) : \citet{Becker91} (5 GHz).
\end{table}

\begin{table}
\begin{center}
\caption{The FRII radio galaxies.}
\begin{tabular}{lcccccccccc}\hline\hline
IAU      &Alternate&Redshift&$b/a$&log$L_{ext}$&$S_{c}$(5~GHz)&ref.&log$R_c$&log$L_{o}$ &ref.\\
  name   &name     & $z$    &   &W Hz$^{-1}$ &mJy           &    &        &W Hz$^{-1}$&    \\
\hline
 0034-014 &3C 15           &0.073&...&25.46&372.8    &1,3& --0.52 & $<$19.74&1\\
 0035-024 &3C 17$^{\star}$  &0.220&...&26.63&727.9    &1,3& --0.48 & 21.64&1\\
 0038+097 &3C 18           &0.188&...&26.41&118.2    &1,3& --1.18 & 21.36&1\\
 0106+729 &3C 33.1$^{\star}$&0.181&...&26.11&19.7     &1,3& --1.69 & 20.94&1\\
 0109+492 &3C 35           &0.067&...&25.06&23.7     &1,3& --1.40 & $<$19.31&1\\
 0218-021 &3C 63           &0.175&...&26.10&18.3     &1,3& --1.74 & 21.04&1\\
 0307+169 &3C 79           &0.256&...&26.62&14.7     &1,3& --2.03 & 21.02&1\\
 0325+023 &3C 88           &0.030&...&24.86&197.2    &1,3& --0.97 & 19.46&1\\
 0356+102 &3C 98           &0.030&...&25.25&11.1     &1,3& --2.61 & $<$18.53&1\\
 0415+379 &3C 111$^{\star}$ &0.049&...&25.86&1155.3   &1,4& --0.77 & 20.99&1\\
 0433+295 &3C 123          &0.218&...&27.55&85.0     &1,3& --2.33 & $<$19.16&1\\
 0453+227 &3C 132          &0.214&...&26.35&33.5     &1,3& --1.56 & $<$19.55&1\\
 0459+252 &3C 133          &0.277&...&26.85&170.8    &1,3& --1.14 & 20.68&1\\
 0511+008 &3C 135          &0.127&...&25.88&5.5      &1,3& --2.31 & 20.15&1\\
 0605+480 &3C 153          &0.277&...&26.68&0.4      &1,3& --3.56 & $<$19.34&1\\
 0640+233 &3C 165          &0.296&...&26.48&8.7      &1,3& --2.01 & 19.95&1\\
 0642+214 &3C 166          &0.245&...&26.15&553.6    &1,3& --0.03 & 20.71&1\\
 0651+542 &3C 171          &0.238&...&26.49&2.5      &1,3& --2.74 & 19.20&1\\
 0734+805 &3C 184.1        &0.118&...&25.87&7.5      &1,3& --2.23 & 20.96&1\\
 0802+243 &3C 192          &0.060&...&25.56&8.5      &1,3& --2.44 & $<$19.94&1\\
 0818+472 &3C 197.1        &0.131&...&25.82&6.8      &1,3& --2.13 & 20.88&1\\
 0819+061 &3C 198          &0.082&...&25.14&$<$1.5   &5,4& $<$--2.50 & 20.85&1\\
 0917+458 &3C 219$^{\star}$ &0.174&...&26.49&68.7     &1,3& --1.56 & 21.39&1\\
 0936+361 &3C 223          &0.137&...&26.04&11.7     &1,3& --2.07 & $<$20.05&1\\
 0938+399 &3C 223.1        &0.108&...&25.65&8.7      &1,3& --2.02 & $<$19.98&1\\
 0945+076 &3C 227$^{\star}$ &0.086&...&25.94&23.5     &1,3& --2.07 & 21.66&1\\
 0958+290 &3C 234$^{\star}$ &0.185&...&26.34&133.6    &1,3& --1.07 & 21.85&1\\
 1003+351 &3C 236          &0.099&...&25.70&191.5    &1,3& --0.80 & $<$19.84&1\\
 1205+341 &   ...         &0.0788&...&24.46&12.5     &2  & --1.21 & $<$19.81&2\\
 1251+278 &3C 277.3       &0.0857&...&25.37&12.4     &2  & --2.05 & 19.49&2\\
 1319+428 &3C 285         &0.079 &...&25.32&7.8      &1,3& --2.00 & 18.47&1\\
 1330+022 &3C 287.1$^{\star}$&0.216&...&26.34&443.8   &1,3& --0.42 & 21.43&1\\
 1420+198 &3C 300         &0.270  &...&26.58&10.1    &1,3& --2.12 & 20.47&1\\
 1441+522 &3C 303$^{\star}$&0.141  &...&25.83&187.6   &1,3& --0.64 & 21.44&1\\
 1519+078 &3C 318.1       &0.046  &...&24.49&3.0     &6,3& --2.05 & $<$18.52&1\\
 1522+546 &3C 319         &0.192  &...&26.04&1.4    &1,3& --2.72 & $<$20.14&1\\
 1545+210 &3C 323.1$^{\star}$&0.264&...&26.44&43.8   &1,3& --1.36 & 22.83&1\\
 1549+202 &3C 326         &0.089 &0.24$^d$&25.19&15.9   &1,3& --1.46 & $<$19.91&1\\
 1559+021 &3C 327         &0.104 &...     &26.18&40.8   &1,3& --1.91 & $<$19.35&1\\
 1615+325 &3C 332$^{\star}$ &0.152 &...&25.93&11.5            &1,3& --1.89 & 21.56&1\\
 1658+471 &3C 349          &0.205 &...&26.33&21.9            &1,3& --1.76 & 20.88&1\\
 1717-009 &3C 353          &0.030 &...&25.94&216.2           &1,3& --2.01 & $<$18.40&1\\
 1726+318 &3C 357          &0.167 &...&26.10&6.5             &1,3& --2.23 & $<$19.66&1\\
 1825+743 &3C 379.1        &0.256 &...&26.32&3.9             &7,3& --2.32 & $<$19.91&1\\
 1832+474 &3C 381          &0.161 &...&26.18&6.9             &1,3& --2.31 & $<$20.28&1\\
 1833+326 &3C 382$^{\star}$ &0.058 &...&25.47&228.1           &1,4& --0.95 & 22.56&1\\
 1842+455 &3C 388          &0.091 &...&25.80&76.5   &1,3& --1.37 & 20.07&1\\
 1845+797 &3C 390.3$^{\star}$&0.056&...&25.74&434.7  &1,3& --0.97 & 21.88&1\\
 1939+605 &3C 401          &0.201 &...&26.41&47.5   &1,3& --1.52 & 20.35&1\\
 1940+505 &3C 402          &0.025 &...&24.39&48.1   &1,3& --1.27 & 19.46&1\\
 1949+023 &3C 403          &0.059 &...&25.54&12.1   &1,3& --2.28 & 19.49&1\\
 2221-023 &3C 445$^{\star}$ &0.057 &...&25.40&382.8  &1,3& --0.66 & 22.11&1\\
 2243+394 &3C 452          &0.081 &0.27$^l$&25.96&152.3&1,3& --1.33 & $<$19.72&1\\
 2309+090 &3C 456          &0.233 &...&26.24&27.8   &1,3& --1.45 & 21.21&1\\
 2318+235 &3C 460          &0.268 &...&26.02&21.4   &1,4& --1.24 & 20.29&1\\\hline
\end{tabular}
\label{tb-FRII}
\end{center}
Superscripts `$d$' and `$l$' for $b/a$ stand for extended dust disks 
and lanes, respectively. There are only two FRIIs for which $b/a$ for 
extended dust features are available \citep{deKoff00}. We do not 
include the FRIIs in our analysis of the extended dusty disks.
Sources with a star are BLRGs. References :
(1) : \citet{Chiaberge02} (5 GHz, F702W filter, except 3C~192 observed 
with F555W);
(2) : \citet{Capetti02} (1.4 GHz);
(3) : \citet{Kuehr79} (5 GHz);
(4) : \citet{Veron98} (5 GHz);
(5) : \citet{Fomalont78} (5 GHz);
(6) : \citet{Slee89} (1.5 GHz);
(7) : \citet{Spangler85} (1.4 GHz).
\end{table}
\begin{table}
\begin{center}
\caption{The BL~Lac objects.}
\begin{tabular}{lcccccccccc}\hline\hline
IAU &Alternate&Redshift&$m_v$&ref.      &log$L_{ext}$&$S_{c}$(5~GHz)&ref.&log$R_c$&log$L_{o}$ \\
name&name     & $z$&&&W Hz$^{-1}$& mJy          &    &        &W Hz$^{-1}$\\
\hline
 0158+003 &  ...   &0.299&17.96  &1     &24.36&8.38  &2 &0.60  &22.61\\
 0219-164 & ...    &0.698&17.0   &9     &25.50&358.0 &5 &0.84  &23.78\\
 0219+428 & 3C 66A &0.444&15.08  &6     &27.32&814.0 &2$^a$ &0.14  &24.13\\
 0257+344 & ...    &0.247&18.53  &1     &23.22&11.78 &2 &1.69  &22.21\\
 0317+185 &   ...  &0.190&18.12  &1     &23.50&9.85  &2 &1.08  &22.13\\
 0323+022 &  ...   &0.147&16.98  &6     &23.20&55.0  &5 &1.21  &22.36\\
 0414+009 & ...    &0.287&17.11  &6     &24.50&67.0  &5 &0.54  &22.91\\
 0454+844 &  ...   &1.34$^{\star}$&17.3&9&24.21&1400.0&4 &$>$3.37$^{\dagger}$&24.29\\
 0521-365 &  ...   &0.055&14.62  &9     &26.12&3124.0&2$^a$ &--0.14 &22.43\\
 0548-322 &  ...   &0.069&16.05  &8     &24.67&80.0  &2$^a$ &--0.08 &22.06\\
 0607+711 &  ...   &0.267&19.60  &1     &24.79&14.08 &2 &0.29  &21.85\\
 0706+592 &  ...   &0.124&18.40  &9     &24.20&65.0  &5 &0.17  &21.64\\
 0735+178 &  ...   &$>$0.424&15.40&9    &23.82&1990.0&4 &$>$3.2&23.96\\
 0737+746 &  ...   &0.315&16.89  &1     &23.85&24.47 &2 &1.64 &23.09\\
 0851+202 & OJ287  &0.306&13.81  &7     &24.21&2217.0&2$^a$ &3.23 &24.29\\
 1101-232 &  ...   &0.186&17.01  &8     &24.40&49.0  &5 &0.41 &22.56\\
 1101+384 & Mrk421 &0.030&13.22  &6     &23.85&520.0 &2$^a$ &0.81 &22.46\\
 1133+704 & Mrk180 &0.044&14.49  &9     &24.31&131.0 &2$^a$ &0.09 &22.29\\
 1218+304 &  ...   &0.130&15.80  &6     &22.79&62.0  &5 &$>$1.34 &22.72\\
 1219+285 & ON 231 &0.102&15.40  &7     &23.07&2058.0&2$^a$ &3.25 &22.67\\
 1221+248 & ...    &0.218&17.65  &1     &23.64&27.85 &2 &1.53 &22.45\\
 1229+645 & ...    &0.164&16.89  &1     &23.73&42.49 &2 &1.34 &22.49\\
 1235+632 & ...    &0.297&18.59  &1     &24.00&13.0  &5 &$>$0.32 &22.35\\
 1400+162 & ...    &0.244&16.74  &9     &26.27&233.0 &2$^a$ &--0.07 &22.91\\
 1402+042 & ...    &0.344&16.88  &8     &24.40&21.43 &2 &0.53  &23.17\\
 1407+599 & ...    &0.495&19.67  &1     &25.76&14.12 &2 &0.08  &22.39\\
 1418+546 & ...    &0.152&15.39  &6     &24.31&1058.0&2$^a$ &2.09  &23.03\\
 1426+427 &...     &0.130&16.40  &9     &23.50&31.0  &5 &$>$1.04&22.48\\
 1443+638 &   ...  &0.299&19.65  &1     &24.63&8.36  &2 &0.33 &21.93\\
 1458+228 &  ...   &0.235&16.79  &1     &23.98&29.0  &2 &1.28 &22.86\\
 1514-241 & AP Lib &0.049&14.97  &7     &23.61&2562.0&2$^a$ &2.18 &22.19\\
 1534+018 &  ...   &0.312&18.70  &1     &25.28&28.84 &2 &0.26 &22.35\\
 1538+149 &4C 14.60&0.605&17.89  &6     &26.94&1337.0&2$^a$ &1.17 &23.29\\
 1552+203 &  ...   &0.222&17.70  &1     &24.65&33.09 &2 &0.61 &22.44\\
 1652+398 & Mrk501 &0.034&14.08  &6     &23.52&1376.0&2$^a$ &1.66 &22.23\\
 1727+502 &  ...   &0.055&16.12  &6     &23.83&175.0 &2$^a$ &0.89 &21.83\\
 1749+096 &  ...   &0.320&17.32  &6     &23.89&744.0 &2$^a$ &$>$3.11 &22.93\\
 1807+698 & 3C 371 &0.050&14.57  &7     &25.04&1350.0&2$^a$ &0.48 &22.37\\
 2143+070 &  ...   &0.237&18.04  &1     &24.99&44.63 &2 &0.46 &22.37\\
 2155-304 &  ...  &0.117&13.31  &6     &25.18&252.00&2$^a$&0.37 &23.62\\
 2200+420 &BL~Lac &0.069&15.42  &7     &23.93&3310.00&2$^a$&2.26 &22.31\\
 2201+044 & ...   &0.028&15.47  &8     &23.70&316.00 &5&0.25 &21.50\\
 2254+074 & ...   &0.190&16.29  &6     &24.51&454.00 &2$^a$&1.73 &22.87\\
 2356-309 & ...   &0.165&17.18  &8     &23.50&42.00  &5&0.90 &22.38\\\hline
\end{tabular}
\label{tb-BLL}
\end{center}
Redshift with a star is from \citet{RectorStocke01} -- 
log$R_c$ $^{\dagger}$ was thus calculated for this new $z$ using 
the log$R_c$ quoted in \citet{VermeulenCohen94}. 
References : 
(1) : \citet{Morris91}; 
(2) : \citet{PerlmanStocke93} ({\it EMSS} XBLs except 2$^a$ which are RBLs, 
1.4 GHz);
(4) : log$R_c$ from \citet{VermeulenCohen94};
(5) : \citet{Laurent-Muehleisen93} ({\it HEAO-1} XBLs, 1.5 GHz);
(6) : \citet{Pica88};
(7) : \citet{Webb88}; 
(8) : \citet{Falomo94}; 
(9) : \citet{PadovaniGiommi95}.
\end{table}

\begin{table}
\begin{center}
\caption{The Radio-loud quasars.}
\begin{tabular}{lccccccccccc}\hline\hline
IAU &Alternate&Redshift&$m_v$&ref.     &log$L_{ext}$&$S_{c}$(5~GHz)&ref.&log$R_c$&log$L_{o}$ \\
name&name     &$z$     &     &         &W Hz$^{-1}$& mJy          &    &        &W Hz$^{-1}$\\
\hline
 0016+731&  ...   &1.781&19.00 &4      &27.37&$>$1500.0 & 1& $>$0.7  &23.88\\
 0106+013& ...    &2.107&18.39 &4      &27.65& 3470.0   & 1& 0.9     &24.29\\
 0153+744&  ...   &2.338&16.00 &4      &26.26& 1510.0   & 1& $>$2.0  &25.35\\
 0212+735& ...    &2.367&20.00 &4      &25.03& 2200.0   & 1& $>$3.4  &23.76\\
 0234+285& ...    &1.207&18.50 &4      &25.68& 1440.0   & 1& 2.1     &23.71\\
 0333+321&NRAO140 &1.259&17.50 &4      &26.44& 2460.0   & 1& 1.6     &24.15\\
 0458-020& ...    &2.286&19.50 &4      &27.17& 1600.0   & 1& 1.1     &23.93\\
 0615+820& ...    &0.710&17.50 &4      &26.38&$>$900.0  & 1& $>$0.8  &23.60\\
 0711+356& ...    &1.620&19.00 &4      &25.96& 1500.0   & 1& $>$2.1  &23.79\\
 0723+679&3C 179  &0.846&18.00 &4      &27.54& 320.0    & 1&--0.68    &23.57\\
 0835+580&3C 205  &1.536&17.62 &4      &27.89& 23.0     & 1&--1.74    &24.29\\
 0836+710& ...    &2.180&16.50 &4      &27.04& 2550.0   & 1& 1.4     &25.08\\
 0839+616& ...    &0.862&17.85 &4      &26.84& 34.0     & 1&--0.94    &23.64\\
 0850+581& ...    &1.322&18.00 &4      &27.45& 1090.0   & 1& 0.27    &23.99\\
 0906+430&3C 216  &0.670&18.10 &4      &27.21& 1060.0   & 1&--0.01    &23.30\\
 0923+392&4C 39.25&0.698&17.86 &4      &26.77& 7320.0   & 1& 1.3     &23.44\\
 1039+811& ...    &1.260&16.50 &4      &26.20& 1120.0   & 1& 1.5     &24.55\\
 1040+123&3C 245  &1.028&17.29 &4      &27.44& 860.0    & 1& 0.0    &24.04\\
 1150+812& ...    &1.250&18.50 &4      &26.50& 1140.0   & 1& 1.2     &23.74\\
 1156+295& ...    &0.729&14.41 &4      &26.32& 810.0    & 1& 0.83    &24.86\\
 1222+216&4C 21.35&0.435&17.50 &4      &26.47& 420.0    & 1&--0.01    &23.14\\
 1226+023&3C 273  &0.158&12.85 &4      &26.70& 39000.0  & 1& 0.9     &24.08\\
 1253-055&3C 279  &0.538&17.75 &4      &27.07& 14500.0  & 1& 1.1     &23.24\\
 1458+718&3C 309.1&0.905&16.78 &4      &27.63& 2680.0   & 1& 0.2     &24.12\\
 1641+399&3C 345  &0.594&15.96 &4      &26.33& 5520.0   & 1& 1.5     &24.05\\
 1642+690&  ...   &0.751&20.50 &4      &26.67& 1260.0   & 1& 0.7     &22.45\\
 1721+343&4C 34.47&0.206&15.46 &4      &25.95& 470.0    & 1&--0.05    &23.27\\
 1828+487&3C 380  &0.691&16.81 &4      &27.32& 6590.0   & 1& 0.7     &23.85\\
 1830+285&4C 28.45&0.594&17.16 &4      &27.02& 450.0    & 1&--0.28    &23.57\\
 1901+319&3C 395  &0.635&17.50 &4      &26.81& 1480.0   & 1& 0.5     &23.49\\
 1928+738&  ...   &0.302&16.06 &4      &25.75& 3210.0   & 1& 1.3     &23.38\\
 1951+498&  ...   &0.466&17.50 &4      &26.09& 91.0     & 1&--0.24    &23.20\\
 2223-052&3C 446  &1.404&18.39 &4      &28.24& 2310.0   & 1&--0.15    &23.90\\
 2251+158&3C 454.3&0.859&16.10 &4      &27.05& 9690.0   & 1& 1.3     &24.34\\
\hline
\end{tabular}
\label{tb-QSR}
\end{center}
References : 
(1) : log$R_c$ from \citet{VermeulenCohen94};
(4) : \citet{Veron98}.
\end{table}
\bibliographystyle{aa}
\bibliography{aap}

\end{document}